\documentclass[floatfix,reprint, superscriptaddress, amsmath,amssymb, aps]{revtex4-2}

\usepackage{pgf}
\usepackage[utf8]{inputenc}
\usepackage{graphicx}
\usepackage{dcolumn}
\usepackage{bm}
\usepackage{lipsum}
\usepackage{braket}
\usepackage{siunitx}
\usepackage[colorlinks=true, linkcolor=blue, citecolor=blue, urlcolor=blue]{hyperref}

\begin{document}

\title{Unveiling the interplay of Mollow physics and perturbed free induction decay by nonlinear optical signals of a dynamically driven two-level system} 

\author{Jan M. Kaspari}

\affiliation{Condensed Matter Theory, TU Dortmund, Otto-Hahn-Straße 4, 44227 Dortmund, Germany}
\author{Thomas K. Bracht}
\affiliation{Condensed Matter Theory, TU Dortmund, Otto-Hahn-Straße 4, 44227 Dortmund, Germany}
\author{Katarina Boos}
\affiliation{Walter Schottky Institut, TUM School of Computation, Information and Technology, and MCQST, Technische Universit\"at M\"unchen, 85748 Garching, Germany}
\author{Sang Kyu Kim}
\affiliation{Walter Schottky Institut, TUM School of Computation, Information and Technology, and MCQST, Technische Universit\"at M\"unchen, 85748 Garching, Germany}
\author{Friedrich Sbresny}
\affiliation{Walter Schottky Institut, TUM School of Computation, Information and Technology, and MCQST, Technische Universit\"at M\"unchen, 85748 Garching, Germany}
\author{Kai M\"uller}
\affiliation{Walter Schottky Institut, TUM School of Computation, Information and Technology, and MCQST, Technische Universit\"at M\"unchen, 85748 Garching, Germany}
\author{Doris E. Reiter}
\affiliation{Condensed Matter Theory, TU Dortmund, Otto-Hahn-Straße 4, 44227 Dortmund, Germany}
\date{\today}

\begin{abstract}
Nonlinear optical signals in optically driven quantum systems can reveal coherences and thereby open up the possibility for manipulation of quantum states. While the limiting cases of ultrafast and continuous-wave excitation have been extensively studied, the time-dynamics of finite pulses bear interesting phenomena. In this paper, we explore the nonlinear optical probe signals of a two-level system excited with a laser pulse of finite duration. In addition to the prominent Mollow peaks, the probe spectra feature several smaller peaks for certain time delays. Similar features have been recently observed for resonance fluorescence signals [arxiv:2305.15827 (2023)].
We discuss that the emergent phenomena can be explained by a combination of Mollow triplet physics and perturbed free induction decay effects, providing an insightful understanding of the underlying physics.
\end{abstract}

\maketitle

\section{\label{sec:level1}Introduction}
Linear and nonlinear optical signals are powerful tools to investigate quantum emitters in order to achieve new ways of manipulation for quantum technologies. In such signals, the quantum emitter is excited by a single or a set of laser pulses and the system's response is monitored with different outputs depending on the signal of interest. 

One class of optical signals is resonance fluorescence, i.e., the emission of a resonantly, coherently driven system \cite{kimble1976theory,muller2007}. A prominent example of coherent physics in resonance fluorescence is the Mollow triplet occuring for driving with a continuous-wave (cw) excitation \cite{mollow1969}. The Mollow triplet has been measured for quantum emitters like semiconductor quantum dots \cite{ulhaq2012,ulhaq2013,gustin2021} or other artificial two-level systems like superconducting circuits \cite{astafiev2010resonance}. Moving from cw excitation to finite pulses, a more complex spectrum emerges in the resonance fluorescence \cite{rodgers1991,moelbjerg2012,konthasinghe2014resonant,florjaczyk1985,buffa1988}. For finite Gaussian pulses, additional peaks appear between the central and outer Mollow lines. Measuring time-dependent dressed-states poses a high demand on the experimental setup and measurement fidelity, especially for quantum emitters working in the optical regime. Therefore, the full emission spectrum of a quantum dot, showing multiple side peaks, was only experimentally detected recently \cite{boos2023signatures, liu2023dynamic}, coined as dynamical Mollow triplet. 

Another class of optical signals are spectra gained in coherent control experiments \cite{axt2004femtosecond}. In such experiments, a pump pulse creates a coherence in the system, which is probed by following pulses. In most cases, the exciting laser pulses are much shorter than the time scales of the system. In the simplest case, two pulses are employed in pump-probe spectroscopy \cite{danckwerts2006theory,sotier2009femtosecond,henzler2021femtosecond,zecherle2010ultrafast}, but also four- or six-wave-mixing spectrocopy techniques are commonly used \cite{suzuki2016coherent,fras2016multi,richter2018deconvolution,grisard2022multiple}. Interestingly, if the order of pulses is inverted, i.e., the probe pulse comes before the pump pulse, a perturbed free induction decay occurs \cite{betz2012,guenther2002,yan2011,nuernberger2009,mondal2018,murotani2018,brosseau2023,hamm1995,seidner1995,wolpert2012}. In the spectra, this manifest as fringes in addition to the main peak.

Motivated by recent experimental findings on the dynamical Mollow spectrum \cite{boos2023signatures, liu2023dynamic}, in this paper, we study the emergence of the observed features by simulating the optical signals of a finite excitation pulse within a pump-probe setup. Choosing a pump-probe configuration over resonance fluorescence has the advantage of access to the dynamics of the system. For this, we extend the methods presented in Refs.~\cite{Reiter2017,klassen2021optical}, where the case of a cw pump pulse was studied, to finite pulses. Due to the knowledge of the time-resolved dynamics, we can unveil the interplay of Mollow triplet physics and perturbed free induced decay in the probe spectra. Comparing our findings in the probe spectra to the spectra measured in resonance fluorescence \cite{boos2023signatures}, we find a good agreement between the main features as the physics behind both signals is the same.  

\section{Theoretical Background}
\begin{figure}
    \centering
        \includegraphics[width=0.48\textwidth]{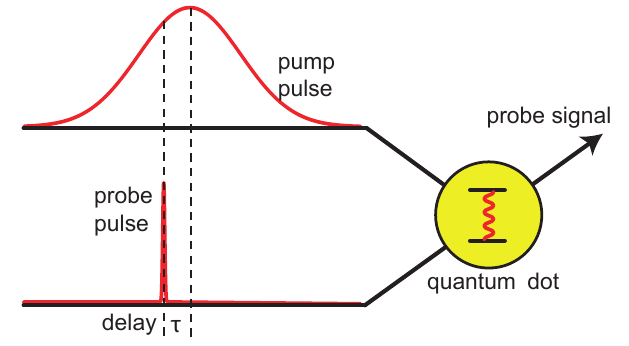}
            \caption{Scheme of a pump-probe set-up as described in detail in the text.}
            \label{fig:scheme:pump-probe}
\end{figure}

To model the optical signals we consider a two-level system, which is a common model for quantum emitters like quantum dots. The two levels are the ground state $\ket{g}$ and the exciton state $\ket{x}$ separated by the energy difference $E_{x}$. The coupling to an external driving light field is described semiclassically in the dipole and rotating wave approximation. The light field energy $\hbar\omega_L$ is assumed to be resonant to the transition between the two states. Under these conditions the Hamiltonian in the rotating frame is given by 
\begin{equation}
    H=-\frac{\hbar}{2}\left(\Omega^\ast(t)\ket{g}\bra{x}+\Omega(t)\ket{x}\bra{g}\right),
\end{equation}
where $\Omega(t)$ denotes the complex, temporal envelope of the light field. Considering a pump-probe setup, we are interested in the probe signals under different pulse shapes of the pump laser. A schematic drawing of such a set-up is shown in Fig. \ref{fig:scheme:pump-probe}. In these kinds of experiments, a pump pulse is used to excite the system, i.e., inducing a dynamic change. After a time delay $\tau$, with respect to the pump pulse, a second pulse is used to probe the response of the system. While the pump pulse can be strong, e.g., it can cause a complete population inversion, the ultrashort probe pulse is so weak that it hardly influences the dynamics of the system. 

The envelope of the complete light field is composed of the pump pulse $\Omega_{\text{pump}}$ and the probe pulse $\Omega_{\text{probe}}$ 
\begin{equation}
    \Omega(t)=\Omega_{\text{pump}}(t)+\Omega_{\text{probe}}(t-\tau)e^{i\Phi}.
\end{equation}
We assume a phase factor $e^{i\Phi}$ between pump and probe pulse. Using the Hamiltonian with the two light fields, we set up the equations of motion and solve these by numerical integration.

To calculate the probe polarization, following Ref. \cite{seidner1995}, we expand the optical polarization $\boldsymbol{P}\left(t,\tau,\Phi\right)$ in a Fourier series with respect to the phase factor
\begin{eqnarray}
    \boldsymbol{P}\left(t,\tau,\Phi\right)=\sum_n\boldsymbol{P}_n\left(t,\tau\right)e^{-in\Phi},
\end{eqnarray}
and perform a phase selection. The desired probe polarization is calculated by the Fourier coefficient 
 \begin{eqnarray} \label{eq:FT}
     \boldsymbol{P}_n\left(t,\tau\right)=\frac{1}{2\pi}\int_0^{2\pi}\,d\Phi\boldsymbol{P}\left(t,\tau,\Phi\right)e^{in\Phi},
 \end{eqnarray}
for $n=1$. 
To obtain the desired absorption spectrum we calculate the imaginary part of the Fourier transformation of the probe polarization \cite{Reiter2017}
\begin{eqnarray}    \alpha\left(\omega\right)=\Im\left[\mathcal{F}\left[P_1\left(t,\tau\right)\cdot e^{-\Gamma t}\right]\right].
\end{eqnarray}
We introduce a damping to the polarization with the damping coefficient $\Gamma=0.139~\text{ps}^{-1}$.
\section{Results}
The results are structured as follows: We begin by briefly summarizing the known limiting cases of the pump pulse being either a cw-pulse switched on instanteneously, or an ultrashort pulse in the $\delta$-pulse limit. Understanding these cases will set the stage to discuss the probe spectra for finite pulses. For this, we start by examining rectangular pulses and then gradually transition to Gaussian pulses by softening the pulse edges. To connect our calculations with real-world numbers, we consider typical time scales of quantum dot dynamics \cite{boos2023signatures}, which gives typical time scale of picoseconds. We conclude by comparing our findings for the probe spectra with measurements in resonance fluorescence from Ref.~\cite{boos2023signatures}.

In all cases, we consider an ultrafast probe pulse, which in the numerical calculations is given by a Gaussian pulse with the envelope 
\begin{equation}
    \Omega_{\text{probe}}(t)=\frac{\alpha_{\text{probe}}}{\sqrt{2\pi \sigma^2_{\text{probe}}}}
      \exp\left({\frac{-(t-\tau)^2}{2 \sigma_{\text{probe}}^2}}\right)
\end{equation}
The pulse area of the probe pulse is set to be very weak with $\alpha_{\text{probe}}=0.02\pi$ and to have an ultrashort pulse we set the pulse width to $\sigma_{\text{probe}}=10$~fs. The delay $\tau$ of the probe pulse defines the time difference between pump and probe pulse as discussed below for the different pump pulses.

\subsection{Limiting cases\label{subsec:cw}} 
The first limiting case is the one of cw-excitation. Thus, for the pump pulse, we consider a constant pulse with pulse strength $\Omega_{\text{cw}}$ switched on instantaneously.
\begin{eqnarray}
    \Omega_{\text{pump}}(t)=\Omega_{\text{cw}}\,\Theta(t-NT_R) 
    \nonumber \, ,
\end{eqnarray}
where $\Theta$ is the Heavy-side function.  We set the starting point of the pulse to $-N$ oscillation periods, such that reference point for the probe pulse the $N$-th maximum of oscillation. This leads to Rabi oscillations with the period $T_R=2\pi/\Omega_{\text{cw}}$.

\begin{figure}    
    \includegraphics[width=\columnwidth]{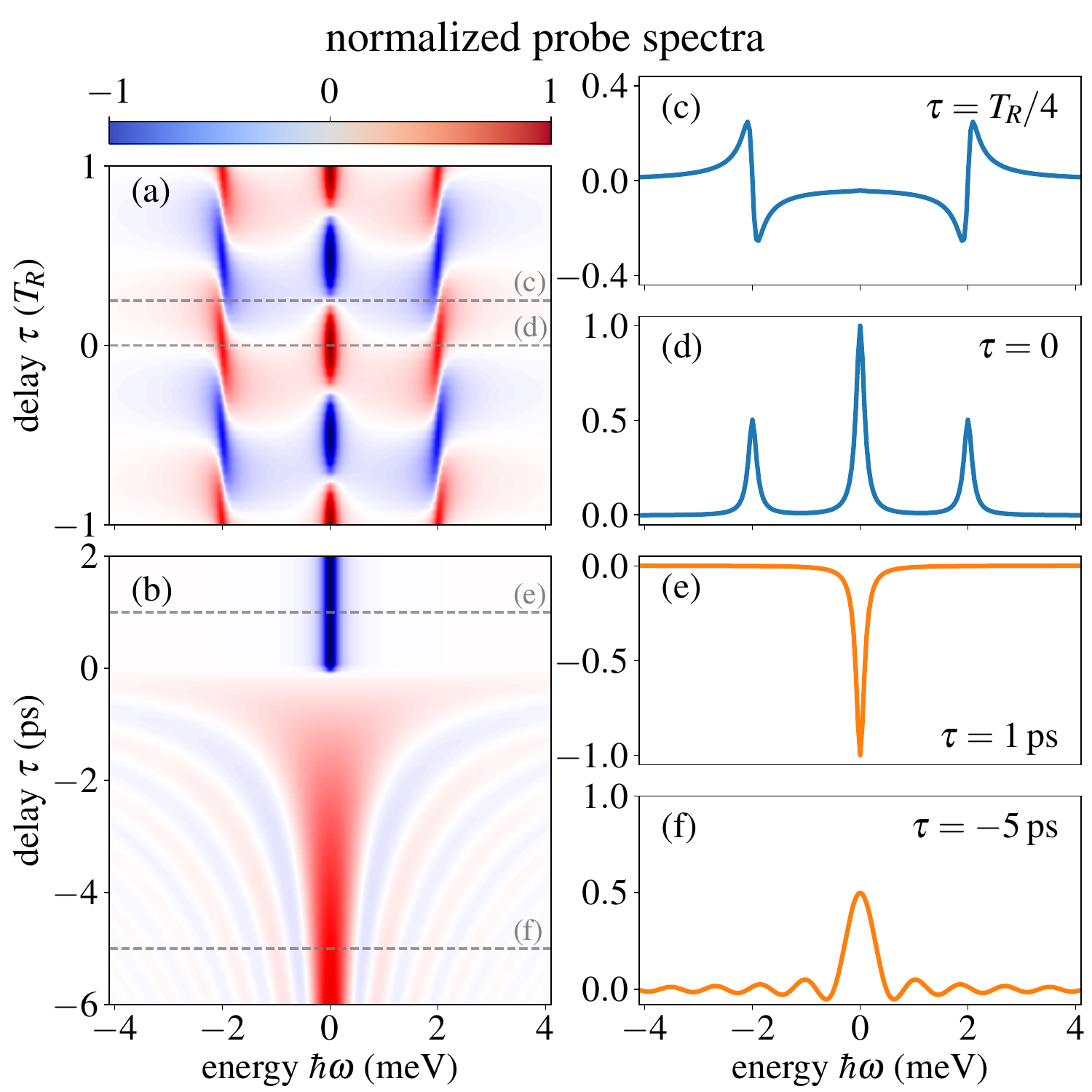}
        \caption{Color plot of the normalized probe spectrum (a) under cw-excitation switched on instantaneously as a function of delay $\tau$ for two periods from $-T_R$ ($-2.06~$ps) to $T_R$ ($2.06~$ps) and (b) for delta-pulse excitation at $\tau=0$ as a function of $\tau$. Cuts showing probe spectra for cw pump at (c) $\tau=T_R/4$ and (d) $\tau=0$ and for ultra-fast excitation at (e) $\tau=1$~ps and (f) $\tau=-5$~ps. }
        \label{fig:2}
\end{figure}

We set $\hbar\Omega_{\text{cw}}=2~$meV, i.e., $\Omega_{\text{cw}}=3.039~$ps$^{-1}$ and $T_R=2.066~$ps. Initially, the system is in its ground state and we choose $N=250$.

The resulting probe spectra for different delays $\tau$ are shown in Fig.~\ref{fig:2}. They are all normalized to the middle peak for $\tau=0$. Due to the periodicity of the Rabi oscillations for a cw-laser, the spectra are the same for $\tau \rightarrow \tau + nT_R$, with $n\in\mathcal{N}$.
Theses spectra exhibit three distinct peaks at $\hbar\omega=0$ and $\hbar\omega=\pm\hbar\Omega_R$, as known from the Mollow triplet. The amplitude variations in these peaks reflect the dynamics of the two-level system.
When the peaks have a positive amplitude, the system is in the ground state, while negative amplitudes correspond to the system being in the excited state. Consequently, we refer to positive peaks as absorption and to negative peaks as gain.
For the time delays $\tau=T_R/4$ and $\tau=3T_R/4$, both states are half occupied resulting in the disappearance of the middle peak, with gain and absorption balancing each other out. Exemplary cuts of the spectra are shown in Fig. \ref{fig:2}(c) for $T_R/4$ and in Fig. \ref{fig:2}(d) for $\tau=0$ . This system dynamics, also including phonons and detuning, is in detail discussed in Ref.s~\cite{Reiter2017,klassen2021optical}.
 
In the second limiting case, we consider an ultrafast pump pulse, such that there is no overlap between pump and probe pulse. The resulting spectra are shown in Fig. \ref{fig:2}(b). In the simulations, we approximate the $\delta$-pulses with ultrashort Gaussian pulses:
\begin{equation}
     \Omega_{\text{pump}}(t)=
        \frac{\alpha_{\text{uf}}}{\sqrt{2\pi \sigma^2_{\text{uf}}}}
        e^ {- \frac{t^2}{2\sigma_{\text{uf}} }} \quad
     \stackrel{\sigma^2_{\text{uf}}\to 0}{\longrightarrow} \quad \Omega_{\text{uf}} \delta(t)
\end{equation}
Here, $\alpha_{\text{uf}}$ is the pulse area of the pulse, while in the numerics we assume the pulse with $\sigma_{\text{uf}}=\sigma_{\text{probe}}$. The time delay refers to the difference between the two pulses. 
 
In most cases, the order of pump and probe pulses is such that the pump pulse excites the system and the probe pulse arrives after the pump pulse. In our model, this corresponds to $\tau>0$. Using a pulse area of $\pi$, the pump pulse inverts the system and the excited state is fully occupied. Accordingly, for $\tau>0$ we see a single negative peak at $\hbar\omega=0$, corresponding to gain.

A distinctly different spectrum emerges for a negative time delay $\tau<0$, where the probe pulse preceeds the pump pulse. Instead of a single peak, we find a strong peak at $\hbar\omega=0$, but with an amplitude smaller than one, next to a series of side peaks. An example is shown in Fig. \ref{fig:2}(f). This is known as perturbed free induction decay \cite{betz2012,guenther2002,yan2011,nuernberger2009,mondal2018,murotani2018,brosseau2023,hamm1995,seidner1995,wolpert2012}. With the probe pulse, the probe polarization starts to oscillate, but is then perturbed by the pump pulse.
This results in a rectangular window for the probe polarization, which in the spectrum corresponds to a sinc-function. The sinc-function is now convoluted with the single peak, which describes the ripple structure observed in Fig.~\ref{fig:2}(b) and (f). We note that the occurance of the ripple structure is strongly connected to the damping time $\Gamma$ used to calculate the polarization (cf. Eq.~\eqref{eq:FT}). If the probe polarization has decayed, before the  pump pulse sets in, the perturbed free induction decay is not observed. 

\subsection{Rectangular pump pulse}

\begin{figure}
    \centering
    \includegraphics[width=\columnwidth]{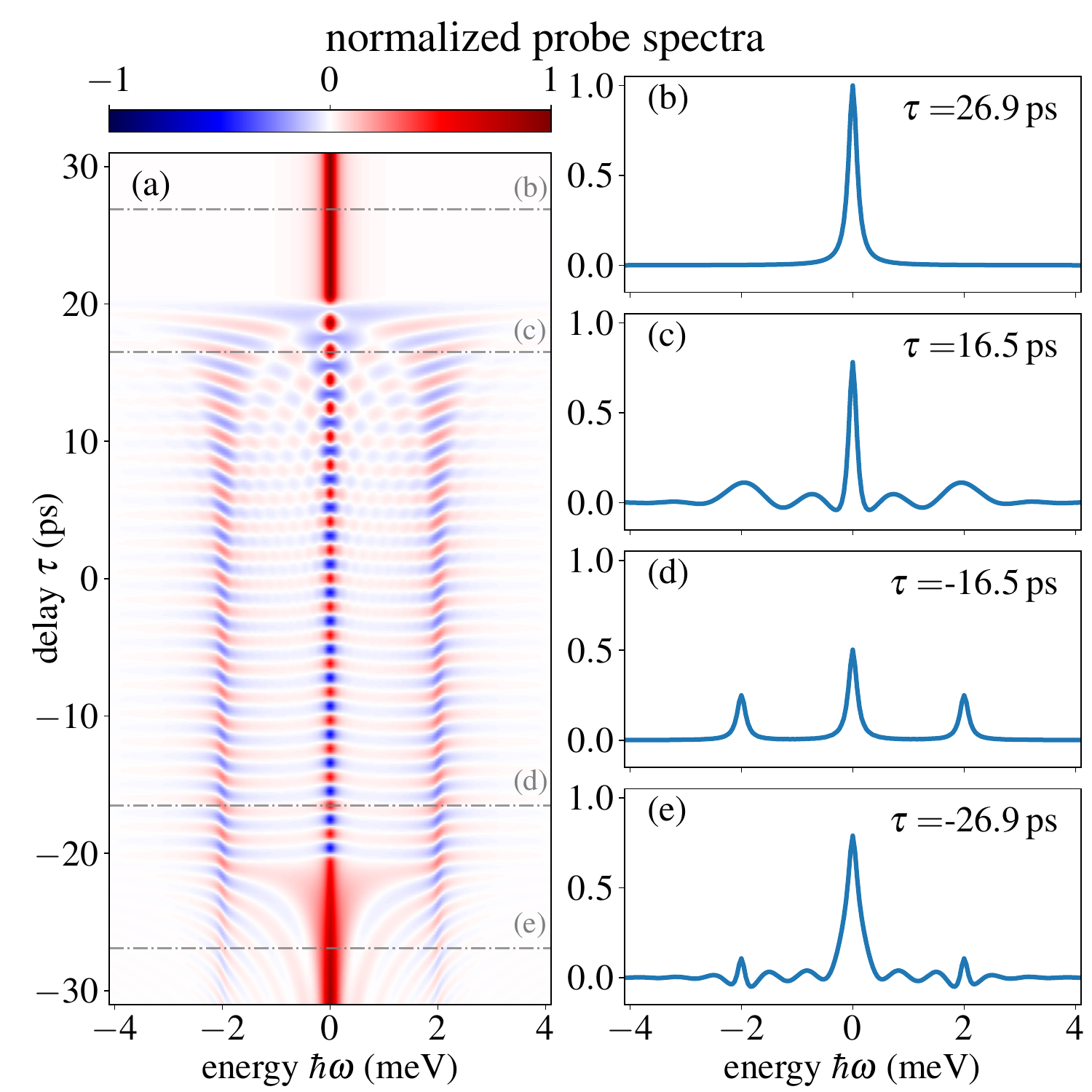}
    \caption{(a) Color plot of the probe spectra for a rectangular pulse acting from $-20.6~$ps ($-10T_R$) to $20.6~$ps ($10T_R$). Cuts through the color plot (highlighted with gray lines) show the probe spectra at (b) $\tau=26.9$~ps, (c) $\tau=16.5$~ps, (d) $\tau=-16.5$~ps and (e) $\tau=-26.9$~ps.}
    \label{fig:3}
\end{figure}

Now we want to expand the model to finite pulses. Coming from the limiting cases, the most simple case is a rectangular pulse of the form

\begin{eqnarray}
    \Omega_{\text{pump}}(t) 
    &=&\Omega_{\text{rect}}\,\text{rect}\left(\frac{t}{NT_R}\right) \nonumber \\
    &=& \Omega_{\text{rect}}\,\Theta\left(t+\frac{1}{2} NT_R\right) \Theta\left(\frac{1}{2} NT_R-t\right)
    \nonumber \ .
\end{eqnarray}
We set $\hbar\Omega_{\text{rect}}=2~$meV, i.e., $\Omega_{\text{rect}}=3.039~$ps$^{-1}$ and $T_R=2.066~$ps. The total pump pulse duration is chosen such that $N=20$ oscillation periods are included in the pulse, which results in a duration of $T_{\text{total}}=NT_R=41.32~$ps. As reference point for the time delay, we choose the center of the rectangluar pulse. 

The probe spectrum for a rectangular pulse as a function of time delay $\tau$ is depicted in Fig. \ref{fig:3}. The pump pulse is centered at $t=0$, thus, $\tau<-20.66~$ps refers to the probe pulse coming before the pump pulse and $\tau>20.66~$ps after the pump pulse. We can discriminate four different regimes depending on the delay $\tau$. We will explain the behaviour using four selected spectra shown in Fig. \ref{fig:3}(b)-(e). The simplest case is shown in Fig. \ref{fig:3}(b), where the probe pulse comes after the pump pulse. After the pump pulse, the system is in its ground state and, in agreement with the limiting case of a $\delta$-pulse, we see a single absorption line at $0~$meV. The other limiting case of a cw-excitation is achieved shortly after the pump pulse is switched on, e.g., at $\tau=-16.5~$ps, as shown in Fig. \ref{fig:3}(d). Here, we again see three absorption lines as typical for a Mollow-type spectrum. 

\begin{figure}[t]
    \includegraphics[width=\columnwidth]{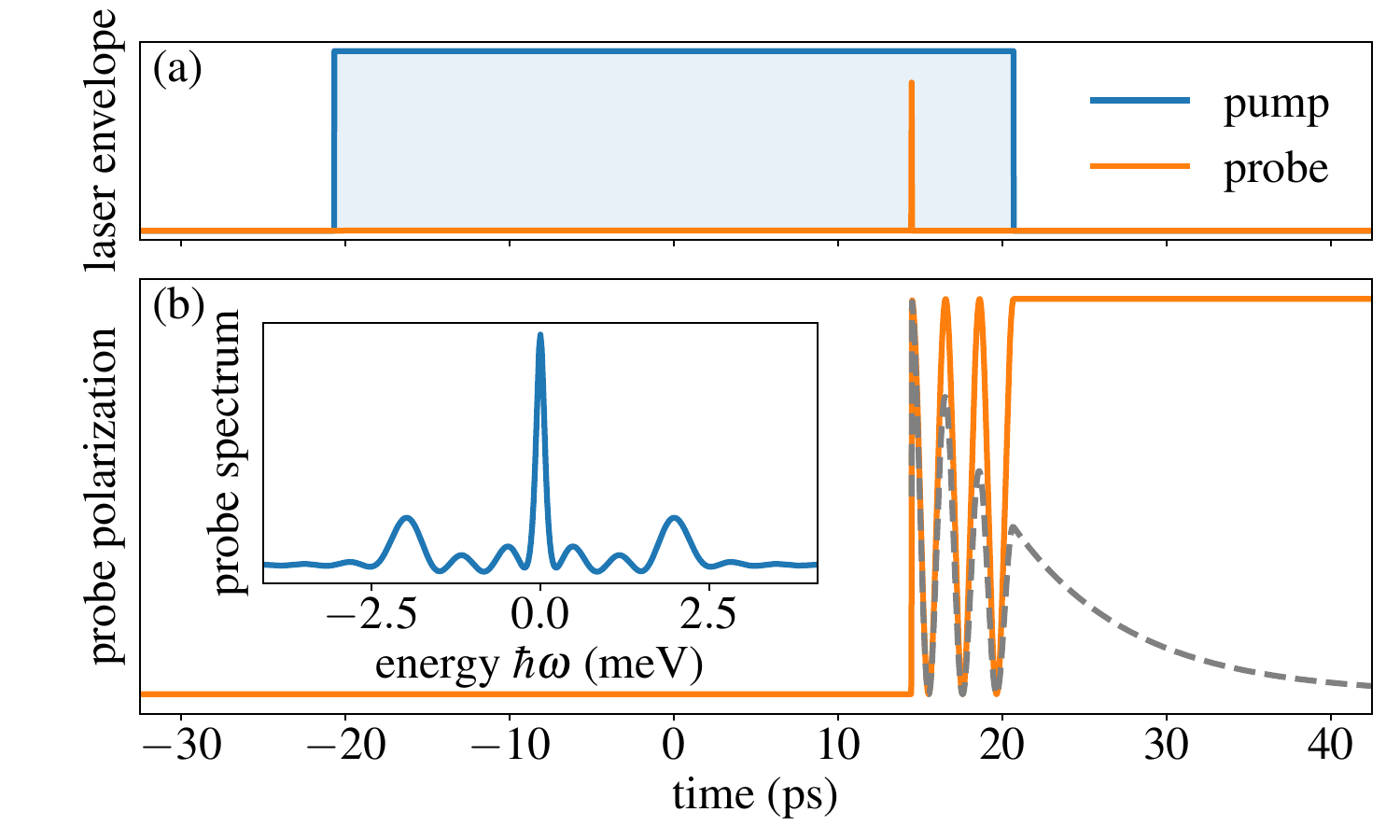}
    \caption{(a) Laser envelopes of a rectangular pump pulse and an ultrashort probe pulse at $\tau=14.47~$ps and corresponding 
    (b) probe polarization (orange), damped probe polarization (gray) and probe spectrum (inset).}
    \label{fig:dynamics_probe}
\end{figure}

A quite different behaviour occurs for probe pulses near the end of the pump pulse as shown in Fig. \ref{fig:3}(c). We observe an additional ripple structure modifying the Mollow triplet. In addition, the amplitude of the Mollow peaks is reduced. This is a consequence of effects similar to the perturbed free induction decay we already observed in the limiting case of ultrashort excitation with a pump pulse preceding the probe pulse. This results in the probe polarization being cut out with a rectangular window, defined by the probe pulse and the ending of the pump pulse. To illustrate this in more detail we show the dynamics of the probe polarization together with the laser pulse sequence in Fig.~\ref{fig:dynamics_probe}. Note that in agreement with the rotating frame we only show the modulation of the probe polarization.
The probe pulse starts the probe dynamics at its onset at $t=14.47$~ps and oscillates with the Rabi frequency during the action of the pump pulse. When the pump pulse ends, the probe polarization becomes constant again. The dashed curve in Fig. \ref{fig:dynamics_probe}(b) shows the damped probe polarization used for the Fourier transform. This behaviour yields a spectrum that is a convolution of the Mollow triplet known from the cw-limit with a sinc-function. 
 
In the last case [Fig.~\ref{fig:3}(f)], when the probe pulse acts before the pump pulse, i.e., $\tau<-20.66~$ps, there is no dynamical pattern and the central peak remains positive. As in the $\delta$-pulse case, a ripple pattern occurs as expected for the perturbed free induction decay. Interestingly, the system already feels the on-set of the cw pulse and a signature of the Mollow triplet is observed. 

\subsection{Smeared out laser pulse}\label{subsec:smeared}

\begin{figure}[t]
    \includegraphics[width=\columnwidth]{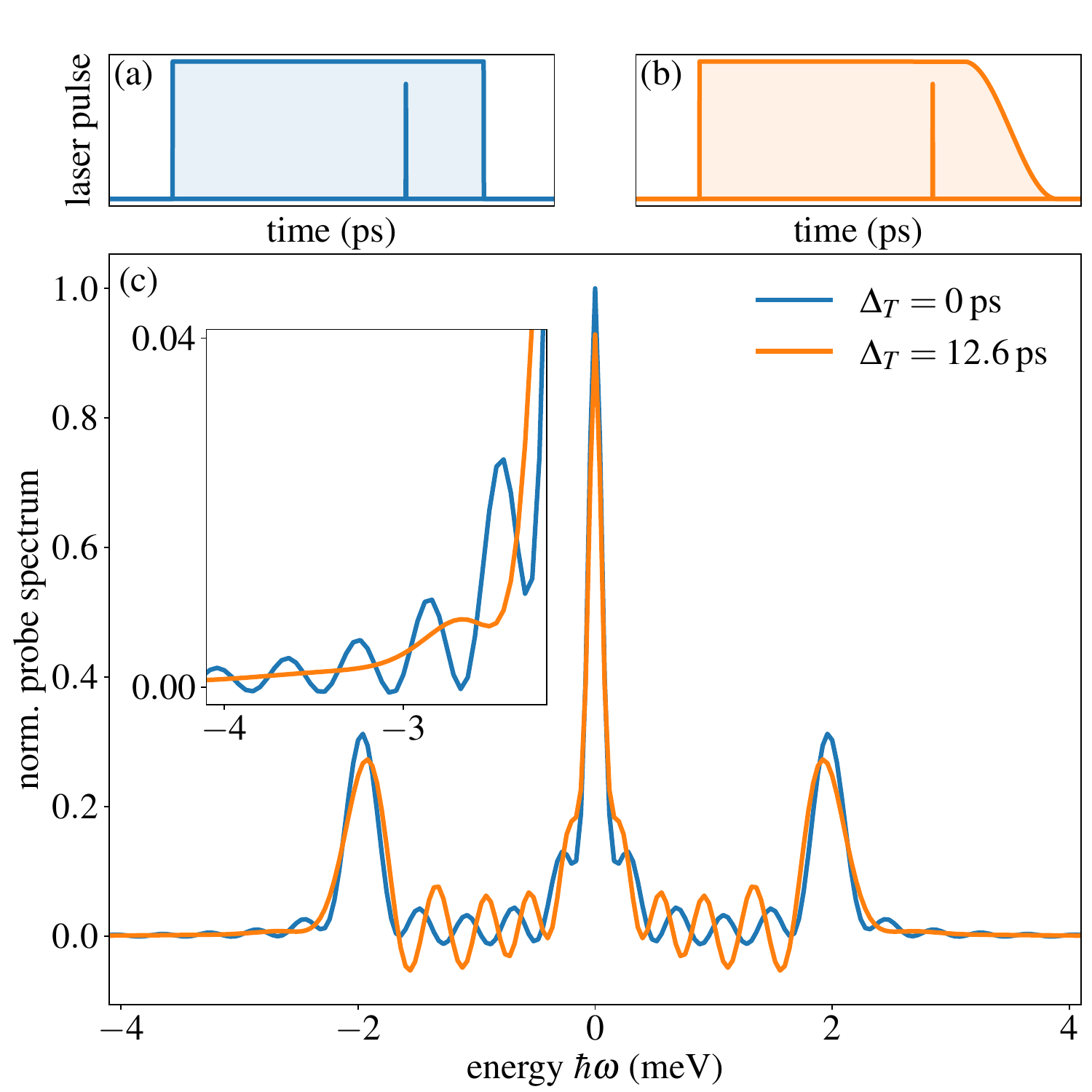}
    \caption{Probe spectra for a rectangular pulse with (blue, $\Delta_T=0$) sharp edge and (orange, $\Delta_T=12.6$~ps) softened edge. Pulse sequences are shown above. In both cases probing is performed $30.9~$ps after the pump pulse has started.}
    \label{fig:4}
\end{figure}

In the next step, we analyse the influence of the pulse shape on the probe spectra during the pulse. We have seen that the switch-off is essential for the ripple pattern observed in addition to the Mollow triplet. Therefore, we soften the switch-off edge of the rectangular pulse, while maintaining a constant pulse area for the whole pulse. To achieve this, we divide our pulse into two separate parts, namely the constant rectangular part and a softened edge

\begin{eqnarray}
    \Omega_{\text{pump}}(t) &=& \Omega_{\text{rect}}(t) + \Omega_{\text{off}}(t) \, .\notag 
\end{eqnarray}
The softened edge is described by a cos$^2$-function with a period of $2\Delta_T$ and a duration of $\Delta_T$ with the equation given in the Appendix. In the case of a sharp edge, the rectangular part is switched on at $t_\text{on}$ and off at $t_\text{off}$. For a soft edge, both the pulses' position and its duration get adjusted in such a way, that the pulse area over the whole pulse stays constant and the start of the pulse does not change. The time delay $\tau$ is chosen to coincide with a maximum of occupation shortly before the end of the pulse, i.e., to be in the same regime as in Fig.~\ref{fig:3}(c). Two examples for pump pulses are depicted in Fig. \ref{fig:4}(a) and (b) for the cases $\Delta_T=0~$ps and $\Delta_T=12.6~$ps respectively. 

The resulting spectra for these two pulses are shown in Fig. \ref{fig:4}(c). Both cases capture a similar structure: They both show a three-peak structure with additional minor peaks inbetween. Our focus lies on the behaviour of these minor peaks. For the rectangular pulse, there are three peaks between each side and the central peak, all with similar strength. In the case of the softened edge, all peaks are slightly shifted to the main center peak. Additionally, their heights vary strongly from each other and increase the further away the associated peak is from the center. Another important difference is found outside of the outer Mollow peaks, which is highlighted in the inset of Fig. \ref{fig:4}(c). We observe that peaks in this area, although already weak in the other cases studied, are damped away almost entirely. 

\subsection{Gaussian laser pulse}
\subsubsection{Pulse delay dependence}
\begin{figure}
    \includegraphics[width=\columnwidth]{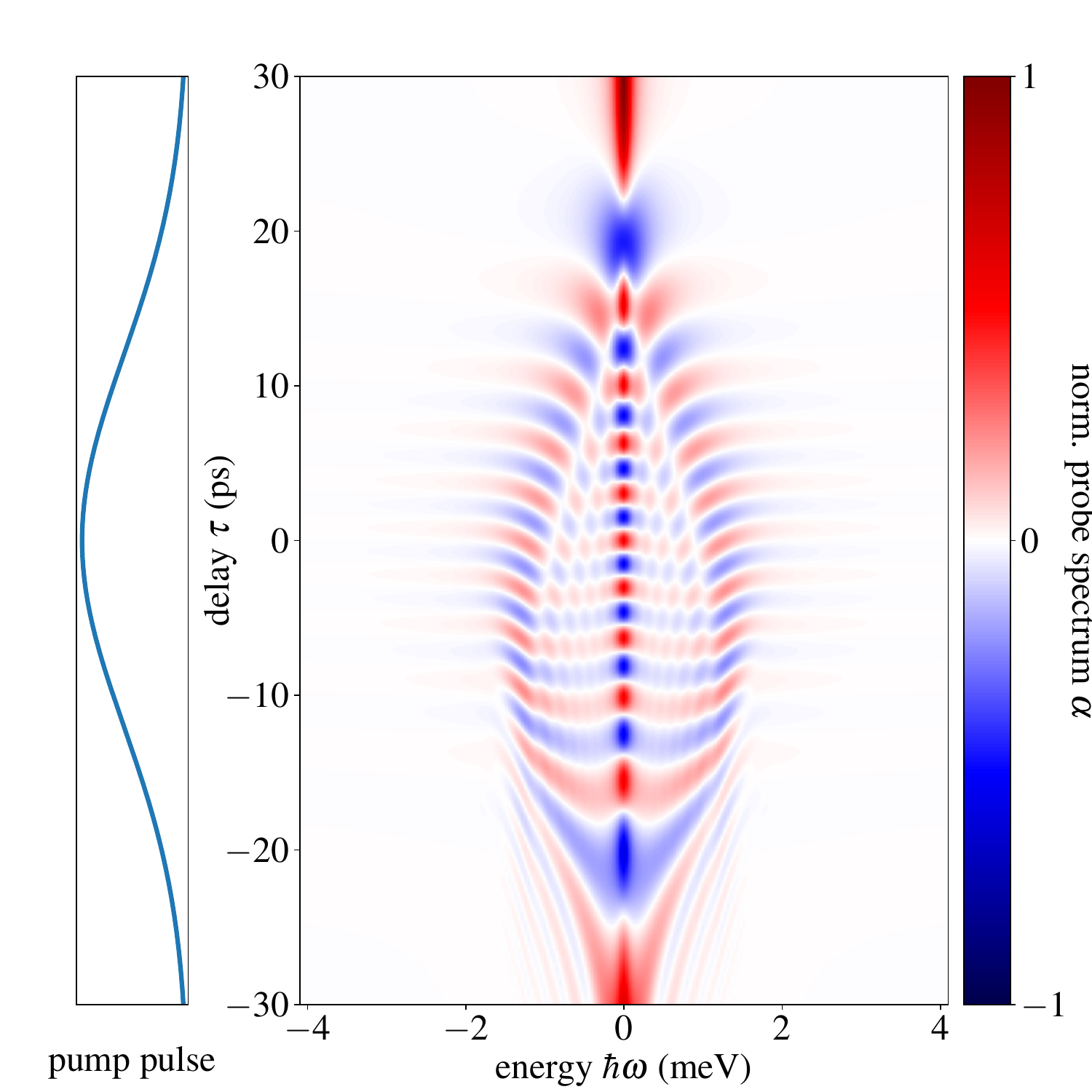}
    \caption{Color plot of the probe spectra during the excitation with a Gaussian pulse for different time delays $\tau$. The temporal profile of the pulse is depicted on the left side.}
    \label{fig:figure6}
\end{figure}
With these findings in mind we want to investigate the spectrum of the probe signal of a Gaussian shaped laser pulse, as used in many experiments 
\begin{equation}
 \Omega_{\text{pump}}(t)=\frac{\alpha}{\sqrt{2\pi\sigma^2_P}}     \exp\left({- \frac{t^2}{2\sigma^2_P}}\right) \,.
\end{equation}
The normalized probe spectra as a function of pulse delay are depicted in Fig.~\ref{fig:figure6}. For the Gaussian pulse, it is less obvious to define delays for which the probe pulse is clearly before or after the pulse. We speak of the onset of the pulse at roughly $-20~$ps, such that a delay of $\tau<-20~$ps corresponds to being before the pulse. Correspondingly, for delays of $\tau>20~$ps the pump pulse is mostly gone, so we speak of after the pulse. Before and after the pump pulse, the spectra are as expected. Of high interest is the behaviour during the pulse for delays $-20~\text{ps}<\tau < 20~$ps. For the Gaussian pulse, the instantaneous Rabi frequency changes continously during the pulse. Accordingly, in the probe spectrum, we can identify two side peaks following the Gaussian shape of the pump pulse as a function of energy. The maximum splitting of the outer peaks is reached for $\tau=0~$ps. In addition, the oscillation in the time delay due to the oscillating occupation is visible as well. At the same time, we notice the manifestation of an additional checkered pattern and a broadening of the peaks towards the end of the Gaussian pulse. All of these observations are consequences of combined effects from Mollow triplet and perturbed free induction decay physics.

\subsubsection{Pulse area dependence}
\begin{figure}
    \includegraphics[width=\columnwidth]{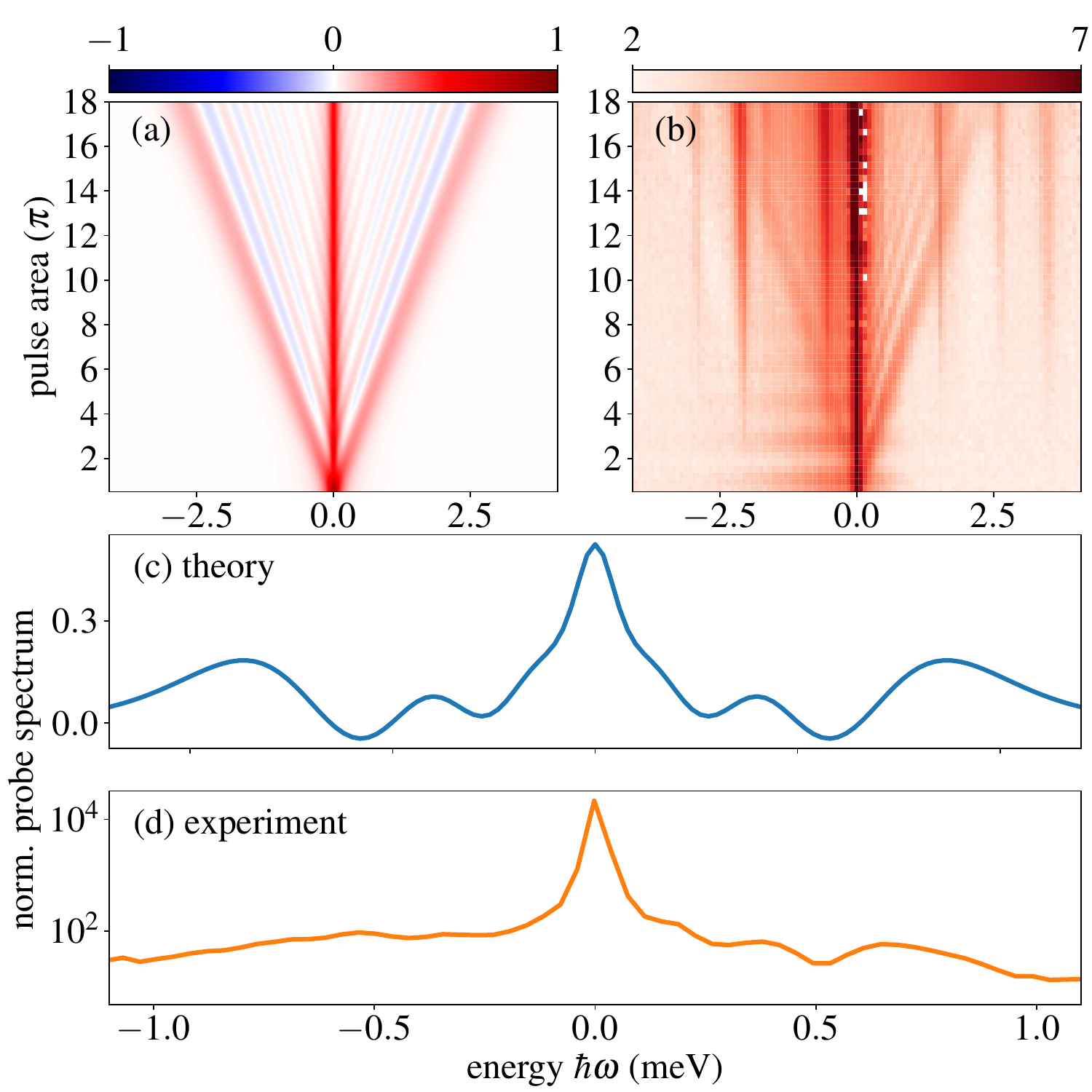}
    \caption{Comparison of (a) theoretical and (b) experimental data  for spectra with Gaussian shaped pulses with a duration (FWHM) of $12~$ps for different pulse areas. Cuts comparing the (c) theoretical and (d) experimental data for a pulse area of $6\pi$. Here, the values for pulse areas and pulse duration of the theoretical pulses refer to after the probe pulse.}
    \label{fig:figure7}
\end{figure}
Now that we understand the general concept of the underlying physics and especially the dependence of the spectrum on the pulse delay, we aim to match our findings with the observations in the resonance fluorescence experiments \cite{boos2023signatures}. These experiments do not have access to the time-resolved dynamics during the pulse, but still the same ripple structure is observed in the spectra. Furthermore, increasing the pulse area results in the emergence of a stripe pattern. In order to match these findings, we investigate how the spectra depend on the pulse area. For that, we construct our pump pulse in such a way that a rectangular shaped pulse part ensures that the system is always in the same state, i.e., the ground state, when the probe pulse hits the system at $t_\text{probe}$. Doing so eliminates the additional amplitude dynamics resulting from the system's state dynamics. Immediately after the probe pulse, the switch-off process begins, where the laser pulse is simulated by half a Gaussian pulse. The equations are given in the appendix. 

We compare our theoretical results for varying pump pulse areas in Fig. \ref{fig:figure7}(a)  with experimental data of the dynamical Mollow spectra in Fig. \ref{fig:figure7}(b). Cuts of the spectra for a pulse area of $6\pi$ are displayed in Fig. \ref{fig:figure7}(c) and (d). We obtain a central peak at $\hbar\omega=0$ for all pulse areas. With increasing pulse area, the effective Rabi frequency increases approximately linearly, which is seen by the energy of the outermost side peaks. Inbetween we find the stripe pattern induced by the finite pulse length. For every pulse area increment of roughly $2\pi$, an additional set of peaks inside the Mollow triplet appears. The same qualitative behaviour is observed in the experimental data. 
The spectra for a given pulse area of $6\pi$ also match well with our simulations as shown in Fig. \ref{fig:figure7}(c) and (d). Note that due to our choice of $\Gamma$ we overestimate the magnitude of the side peaks, while for resonance fluorescence, due to the nature of the signal, effects are smaller and a logarithmic scale was used. Besides the side peak, for $\hbar\omega>0$, there are three signatures in both experiment and theory, the uttermost being the Mollow peak and two features from the perturbed free induction decay. On the lower energy $\hbar\omega<0$ side, the theoretical spectrum shows again three peaks, while in the experimental data a more uniform spectrum appears. Here, the interaction with acoustic phonons leads to the prominent phonon sidebands \cite{Besombes2001,Krummheuer2002}. While the phonon relaxation can be described within a rate equation in the dressed states \cite{Reiter2017,klassen2021optical}, phonon sidebands are a result of the polaron formation captured by non-Markovian approaches \cite{krugel2005,roy2012polaron,cosacchi2018path,cygorek2022}. As the phonon sidebands are not the main focus of the work, we refrained from taking up the numerical effort to include these and describe well-established phenomena, but focussed on the ripple structure visible in both pump-probe simulations and resonance fluoresence measurement. 

\section{Conclusion}
We have studied the time-resolved probe spectra of dynamically driven two-level system. For the limiting cases of ultrashort laser pulses and continuous driving, we observed the well-established perturbed free induction decay and the Mollow triplet. We showed, that for rectangular laser pulses, four distinctly different regimes can be identified -  one where the spectral behaviour is dominated by either the Mollow triplet physics and one where the perturbed free induction decay determines the spectrum, and two regimes where the probe spectrum results from the interplay of both effects. In the latter regimes alongside the Mollow triplet an additional ripple pattern appears. Softening the switch-off process of these pulses and therefore changing their shape led not only to shifts of all peaks towards the center and changes in peak heights, but also to the disappearance of the spectrum outside of the outer Mollow peaks. All of the observed spectral features also occur for Gaussian shaped pulses and provide an insightful understanding of the formation and the underlying physics of the recently measured dynamical Mollow spectrum. In conclusion, we found that the probe spectra of nonlinear optical signals for finite pulses depend on both the pump pulse shape and duration, as well as the probe timing. They emerge as a consequence of the interplay between Mollow triplet physics and perturbed free induction decay.

\section*{Acknowledgement}
KB, SKK, FS and KM gratefully acknowledge financial support from the German Federal Ministry of Education and Research via the funding program Photonics Research Germany (Contract No. 13N14846) and the Deutsche Forschungsgemeinschaft (DFG, German Research Foundation) via projects MU 4215/4-1 (CNLG) and Germany's Excellence Strategy (MCQST, EXC-2111, 390814868).

\appendix

\section{Equations for the pulses}
For the pulse with softened edges we use the following expressions
\begin{eqnarray}
    \Omega_\text{rect}(t) &=& \Omega_\text{rect} \cdot \text{rect}\left(\frac{t+\Delta_T/4}{t_\text{off}-\Delta_T/2-t_\text{on}}\right) \,, \notag \\
    \Omega_{\text{off}}(t) &=&\Omega_\text{rect} \cdot \cos^2\left(\frac{\pi}{2\Delta_T} \cdot\left(t-(t_\text{off}-\Delta_T/2)\right) \right) \,. \notag 
\end{eqnarray}

For the pulse area dependence the equation for the pulse reads
\[
    \Omega_{\text{pump}}(t,\alpha)= 
\begin{cases}
    \Omega_{\text{area}}(\alpha) \cdot \text{rect}\left( \frac{t-\frac{t_{\text{probe}}+t_{\text{on}}(\alpha)}{2}}{t_{\text{probe}}-t_{\text{on}}(\alpha)} \right),&  t < t_{\text{probe}}\\
    \frac{2\alpha}{\sqrt{2\pi\sigma^2_P}}     e^{- \frac{(t-t_{\text{probe}})^2}{2\sigma^2_P}},              & t \geq t_{\text{probe}}
\end{cases},
\]

where $\alpha$\, refers to the pulse area after the probe pulse. A smooth transition between the two pulse parts and the desired selective state preparation at $t_{\text{probe}}$, removing additional amplitude dynamics in the probe spectrum, are both achieved by varying $t_{\text{on}}$ and $\Omega_{\text{area}}$ for every value of $\alpha$, such that they read
\begin{eqnarray}
    t_\text{on}(\alpha) &=& t_{\text{probe}}-\frac{\alpha_{\text{prep}}}{\Omega_{\text{area}}(\alpha)}\,, \notag \\
    \Omega_{\text{area}}(\alpha) &=& \frac{\alpha}{\sqrt{2\pi\sigma^2_P}}\,. \notag 
\end{eqnarray}
The pulse area $\alpha_{\text{prep}}$ is chosen in such a way that the system is always in its ground state when its probed i.e., $\alpha_{\text{prep}}=2\pi$ in our calculations. 

\bibliography{bibliography}

\providecommand{\noopsort}[1]{}\providecommand{\singleletter}[1]{#1}%
\begin{thebibliography}{41}%
\makeatletter
\providecommand \@ifxundefined [1]{%
 \@ifx{#1\undefined}
}%
\providecommand \@ifnum [1]{%
 \ifnum #1\expandafter \@firstoftwo
 \else \expandafter \@secondoftwo
 \fi
}%
\providecommand \@ifx [1]{%
 \ifx #1\expandafter \@firstoftwo
 \else \expandafter \@secondoftwo
 \fi
}%
\providecommand \natexlab [1]{#1}%
\providecommand \enquote  [1]{``#1''}%
\providecommand \bibnamefont  [1]{#1}%
\providecommand \bibfnamefont [1]{#1}%
\providecommand \citenamefont [1]{#1}%
\providecommand \href@noop [0]{\@secondoftwo}%
\providecommand \href [0]{\begingroup \@sanitize@url \@href}%
\providecommand \@href[1]{\@@startlink{#1}\@@href}%
\providecommand \@@href[1]{\endgroup#1\@@endlink}%
\providecommand \@sanitize@url [0]{\catcode `\\12\catcode `\$12\catcode
  `\&12\catcode `\#12\catcode `\^12\catcode `\_12\catcode `\%12\relax}%
\providecommand \@@startlink[1]{}%
\providecommand \@@endlink[0]{}%
\providecommand \url  [0]{\begingroup\@sanitize@url \@url }%
\providecommand \@url [1]{\endgroup\@href {#1}{\urlprefix }}%
\providecommand \urlprefix  [0]{URL }%
\providecommand \Eprint [0]{\href }%
\providecommand \doibase [0]{https://doi.org/}%
\providecommand \selectlanguage [0]{\@gobble}%
\providecommand \bibinfo  [0]{\@secondoftwo}%
\providecommand \bibfield  [0]{\@secondoftwo}%
\providecommand \translation [1]{[#1]}%
\providecommand \BibitemOpen [0]{}%
\providecommand \bibitemStop [0]{}%
\providecommand \bibitemNoStop [0]{.\EOS\space}%
\providecommand \EOS [0]{\spacefactor3000\relax}%
\providecommand \BibitemShut  [1]{\csname bibitem#1\endcsname}%
\let\auto@bib@innerbib\@empty
\bibitem [{\citenamefont {Kimble}\ and\ \citenamefont
  {Mandel}(1976)}]{kimble1976theory}%
  \BibitemOpen
  \bibfield  {author} {\bibinfo {author} {\bibfnamefont {H.~J.}\ \bibnamefont
  {Kimble}}\ and\ \bibinfo {author} {\bibfnamefont {L.}~\bibnamefont
  {Mandel}},\ }\bibfield  {title} {\bibinfo {title} {Theory of resonance
  fluorescence},\ }\href {https://doi.org/10.1103/PhysRevA.13.2123} {\bibfield
  {journal} {\bibinfo  {journal} {Phys. Rev. A}\ }\textbf {\bibinfo {volume}
  {13}},\ \bibinfo {pages} {2123} (\bibinfo {year} {1976})}\BibitemShut
  {NoStop}%
\bibitem [{\citenamefont {Muller}\ \emph {et~al.}(2007)\citenamefont {Muller},
  \citenamefont {Flagg}, \citenamefont {Bianucci}, \citenamefont {Wang},
  \citenamefont {Deppe}, \citenamefont {Ma}, \citenamefont {Zhang},
  \citenamefont {Salamo}, \citenamefont {Xiao},\ and\ \citenamefont
  {Shih}}]{muller2007}%
  \BibitemOpen
  \bibfield  {author} {\bibinfo {author} {\bibfnamefont {A.}~\bibnamefont
  {Muller}}, \bibinfo {author} {\bibfnamefont {E.~B.}\ \bibnamefont {Flagg}},
  \bibinfo {author} {\bibfnamefont {P.}~\bibnamefont {Bianucci}}, \bibinfo
  {author} {\bibfnamefont {X.~Y.}\ \bibnamefont {Wang}}, \bibinfo {author}
  {\bibfnamefont {D.~G.}\ \bibnamefont {Deppe}}, \bibinfo {author}
  {\bibfnamefont {W.}~\bibnamefont {Ma}}, \bibinfo {author} {\bibfnamefont
  {J.}~\bibnamefont {Zhang}}, \bibinfo {author} {\bibfnamefont {G.~J.}\
  \bibnamefont {Salamo}}, \bibinfo {author} {\bibfnamefont {M.}~\bibnamefont
  {Xiao}},\ and\ \bibinfo {author} {\bibfnamefont {C.~K.}\ \bibnamefont
  {Shih}},\ }\bibfield  {title} {\bibinfo {title} {Resonance fluorescence from
  a coherently driven semiconductor quantum dot in a cavity},\ }\href
  {https://doi.org/10.1103/PhysRevLett.99.187402} {\bibfield  {journal}
  {\bibinfo  {journal} {Phys. Rev. Lett.}\ }\textbf {\bibinfo {volume} {99}},\
  \bibinfo {pages} {187402} (\bibinfo {year} {2007})}\BibitemShut {NoStop}%
\bibitem [{\citenamefont {Mollow}(1969)}]{mollow1969}%
  \BibitemOpen
  \bibfield  {author} {\bibinfo {author} {\bibfnamefont {B.~R.}\ \bibnamefont
  {Mollow}},\ }\bibfield  {title} {\bibinfo {title} {Power spectrum of light
  scattered by two-level systems},\ }\href
  {https://doi.org/10.1103/PhysRev.188.1969} {\bibfield  {journal} {\bibinfo
  {journal} {Phys. Rev.}\ }\textbf {\bibinfo {volume} {188}},\ \bibinfo {pages}
  {1969} (\bibinfo {year} {1969})}\BibitemShut {NoStop}%
\bibitem [{\citenamefont {Ulhaq}\ \emph {et~al.}(2012)\citenamefont {Ulhaq},
  \citenamefont {Weiler}, \citenamefont {Ulrich}, \citenamefont {Ro{\ss}bach},
  \citenamefont {Jetter},\ and\ \citenamefont {Michler}}]{ulhaq2012}%
  \BibitemOpen
  \bibfield  {author} {\bibinfo {author} {\bibfnamefont {A.}~\bibnamefont
  {Ulhaq}}, \bibinfo {author} {\bibfnamefont {S.}~\bibnamefont {Weiler}},
  \bibinfo {author} {\bibfnamefont {S.~M.}\ \bibnamefont {Ulrich}}, \bibinfo
  {author} {\bibfnamefont {R.}~\bibnamefont {Ro{\ss}bach}}, \bibinfo {author}
  {\bibfnamefont {M.}~\bibnamefont {Jetter}},\ and\ \bibinfo {author}
  {\bibfnamefont {P.}~\bibnamefont {Michler}},\ }\bibfield  {title} {\bibinfo
  {title} {Cascaded single-photon emission from the {Mollow} triplet sidebands
  of a quantum dot},\ }\href {https://doi.org/10.1038/nphoton.2012.23}
  {\bibfield  {journal} {\bibinfo  {journal} {Nat. Photonics}\ }\textbf
  {\bibinfo {volume} {6}},\ \bibinfo {pages} {238} (\bibinfo {year}
  {2012})}\BibitemShut {NoStop}%
\bibitem [{\citenamefont {Ulhaq}\ \emph {et~al.}(2013)\citenamefont {Ulhaq},
  \citenamefont {Weiler}, \citenamefont {Roy}, \citenamefont {Ulrich},
  \citenamefont {Jetter}, \citenamefont {Hughes},\ and\ \citenamefont
  {Michler}}]{ulhaq2013}%
  \BibitemOpen
  \bibfield  {author} {\bibinfo {author} {\bibfnamefont {A.}~\bibnamefont
  {Ulhaq}}, \bibinfo {author} {\bibfnamefont {S.}~\bibnamefont {Weiler}},
  \bibinfo {author} {\bibfnamefont {C.}~\bibnamefont {Roy}}, \bibinfo {author}
  {\bibfnamefont {S.~M.}\ \bibnamefont {Ulrich}}, \bibinfo {author}
  {\bibfnamefont {M.}~\bibnamefont {Jetter}}, \bibinfo {author} {\bibfnamefont
  {S.}~\bibnamefont {Hughes}},\ and\ \bibinfo {author} {\bibfnamefont
  {P.}~\bibnamefont {Michler}},\ }\bibfield  {title} {\bibinfo {title}
  {Detuning-dependent {Mollow} triplet of a coherently-driven single quantum
  dot},\ }\href {https://doi.org/10.1364/OE.21.004382} {\bibfield  {journal}
  {\bibinfo  {journal} {Opt. Express}\ }\textbf {\bibinfo {volume} {21}},\
  \bibinfo {pages} {4382} (\bibinfo {year} {2013})}\BibitemShut {NoStop}%
\bibitem [{\citenamefont {Gustin}\ \emph {et~al.}(2021)\citenamefont {Gustin},
  \citenamefont {Hanschke}, \citenamefont {Boos}, \citenamefont {M\"uller},
  \citenamefont {Kremser}, \citenamefont {Finley}, \citenamefont {Hughes},\
  and\ \citenamefont {M\"uller}}]{gustin2021}%
  \BibitemOpen
  \bibfield  {author} {\bibinfo {author} {\bibfnamefont {C.}~\bibnamefont
  {Gustin}}, \bibinfo {author} {\bibfnamefont {L.}~\bibnamefont {Hanschke}},
  \bibinfo {author} {\bibfnamefont {K.}~\bibnamefont {Boos}}, \bibinfo {author}
  {\bibfnamefont {J.~R.~A.}\ \bibnamefont {M\"uller}}, \bibinfo {author}
  {\bibfnamefont {M.}~\bibnamefont {Kremser}}, \bibinfo {author} {\bibfnamefont
  {J.~J.}\ \bibnamefont {Finley}}, \bibinfo {author} {\bibfnamefont
  {S.}~\bibnamefont {Hughes}},\ and\ \bibinfo {author} {\bibfnamefont
  {K.}~\bibnamefont {M\"uller}},\ }\bibfield  {title} {\bibinfo {title}
  {High-resolution spectroscopy of a quantum dot driven bichromatically by two
  strong coherent fields},\ }\href
  {https://doi.org/10.1103/PhysRevResearch.3.013044} {\bibfield  {journal}
  {\bibinfo  {journal} {Phys. Rev. Res.}\ }\textbf {\bibinfo {volume} {3}},\
  \bibinfo {pages} {013044} (\bibinfo {year} {2021})}\BibitemShut {NoStop}%
\bibitem [{\citenamefont {Astafiev}\ \emph {et~al.}(2010)\citenamefont
  {Astafiev}, \citenamefont {Zagoskin}, \citenamefont {Abdumalikov~Jr},
  \citenamefont {Pashkin}, \citenamefont {Yamamoto}, \citenamefont {Inomata},
  \citenamefont {Nakamura},\ and\ \citenamefont
  {Tsai}}]{astafiev2010resonance}%
  \BibitemOpen
  \bibfield  {author} {\bibinfo {author} {\bibfnamefont {O.}~\bibnamefont
  {Astafiev}}, \bibinfo {author} {\bibfnamefont {A.~M.}\ \bibnamefont
  {Zagoskin}}, \bibinfo {author} {\bibfnamefont {A.}~\bibnamefont
  {Abdumalikov~Jr}}, \bibinfo {author} {\bibfnamefont {Y.~A.}\ \bibnamefont
  {Pashkin}}, \bibinfo {author} {\bibfnamefont {T.}~\bibnamefont {Yamamoto}},
  \bibinfo {author} {\bibfnamefont {K.}~\bibnamefont {Inomata}}, \bibinfo
  {author} {\bibfnamefont {Y.}~\bibnamefont {Nakamura}},\ and\ \bibinfo
  {author} {\bibfnamefont {J.~S.}\ \bibnamefont {Tsai}},\ }\bibfield  {title}
  {\bibinfo {title} {Resonance fluorescence of a single artificial atom},\
  }\href {https://doi.org/10.1126/science.1181918} {\bibfield  {journal}
  {\bibinfo  {journal} {Science}\ }\textbf {\bibinfo {volume} {327}},\ \bibinfo
  {pages} {840} (\bibinfo {year} {2010})}\BibitemShut {NoStop}%
\bibitem [{\citenamefont {Rodgers}\ and\ \citenamefont
  {Swain}(1991)}]{rodgers1991}%
  \BibitemOpen
  \bibfield  {author} {\bibinfo {author} {\bibfnamefont {P.}~\bibnamefont
  {Rodgers}}\ and\ \bibinfo {author} {\bibfnamefont {S.}~\bibnamefont
  {Swain}},\ }\bibfield  {title} {\bibinfo {title} {Multi-peaked resonance
  fluorescence spectra with rectangular laser pulses},\ }\href
  {https://doi.org/https://doi.org/10.1016/0030-4018(91)90618-N} {\bibfield
  {journal} {\bibinfo  {journal} {Opt. Commun.}\ }\textbf {\bibinfo {volume}
  {81}},\ \bibinfo {pages} {291} (\bibinfo {year} {1991})}\BibitemShut
  {NoStop}%
\bibitem [{\citenamefont {Moelbjerg}\ \emph {et~al.}(2012)\citenamefont
  {Moelbjerg}, \citenamefont {Kaer}, \citenamefont {Lorke},\ and\ \citenamefont
  {M\o{}rk}}]{moelbjerg2012}%
  \BibitemOpen
  \bibfield  {author} {\bibinfo {author} {\bibfnamefont {A.}~\bibnamefont
  {Moelbjerg}}, \bibinfo {author} {\bibfnamefont {P.}~\bibnamefont {Kaer}},
  \bibinfo {author} {\bibfnamefont {M.}~\bibnamefont {Lorke}},\ and\ \bibinfo
  {author} {\bibfnamefont {J.}~\bibnamefont {M\o{}rk}},\ }\bibfield  {title}
  {\bibinfo {title} {Resonance fluorescence from semiconductor quantum dots:
  Beyond the {Mollow} triplet},\ }\href
  {https://doi.org/10.1103/PhysRevLett.108.017401} {\bibfield  {journal}
  {\bibinfo  {journal} {Phys. Rev. Lett.}\ }\textbf {\bibinfo {volume} {108}},\
  \bibinfo {pages} {017401} (\bibinfo {year} {2012})}\BibitemShut {NoStop}%
\bibitem [{\citenamefont {Konthasinghe}\ \emph {et~al.}(2014)\citenamefont
  {Konthasinghe}, \citenamefont {Peiris},\ and\ \citenamefont
  {Muller}}]{konthasinghe2014resonant}%
  \BibitemOpen
  \bibfield  {author} {\bibinfo {author} {\bibfnamefont {K.}~\bibnamefont
  {Konthasinghe}}, \bibinfo {author} {\bibfnamefont {M.}~\bibnamefont
  {Peiris}},\ and\ \bibinfo {author} {\bibfnamefont {A.}~\bibnamefont
  {Muller}},\ }\bibfield  {title} {\bibinfo {title} {Resonant light scattering
  of a laser frequency comb by a quantum dot},\ }\href
  {https://doi.org/10.1103/PhysRevA.90.023810} {\bibfield  {journal} {\bibinfo
  {journal} {Phys. Rev. A}\ }\textbf {\bibinfo {volume} {90}},\ \bibinfo
  {pages} {023810} (\bibinfo {year} {2014})}\BibitemShut {NoStop}%
\bibitem [{\citenamefont {Florjaczyk}\ \emph {et~al.}(1985)\citenamefont
  {Florjaczyk}, \citenamefont {Rzazewsk},\ and\ \citenamefont
  {Zakrzewski}}]{florjaczyk1985}%
  \BibitemOpen
  \bibfield  {author} {\bibinfo {author} {\bibfnamefont {M.}~\bibnamefont
  {Florjaczyk}}, \bibinfo {author} {\bibfnamefont {K.}~\bibnamefont
  {Rzazewsk}},\ and\ \bibinfo {author} {\bibfnamefont {J.}~\bibnamefont
  {Zakrzewski}},\ }\bibfield  {title} {\bibinfo {title} {Resonance scattering
  of a short laser pulse on a two-level system: Time-dependent approach},\
  }\href {https://doi.org/10.1103/PhysRevA.31.1558} {\bibfield  {journal}
  {\bibinfo  {journal} {Phys. Rev. A}\ }\textbf {\bibinfo {volume} {31}},\
  \bibinfo {pages} {1558} (\bibinfo {year} {1985})}\BibitemShut {NoStop}%
\bibitem [{\citenamefont {Buffa}\ \emph {et~al.}(1988)\citenamefont {Buffa},
  \citenamefont {Cavalieri}, \citenamefont {Fini},\ and\ \citenamefont
  {Matera}}]{buffa1988}%
  \BibitemOpen
  \bibfield  {author} {\bibinfo {author} {\bibfnamefont {R.}~\bibnamefont
  {Buffa}}, \bibinfo {author} {\bibfnamefont {S.}~\bibnamefont {Cavalieri}},
  \bibinfo {author} {\bibfnamefont {L.}~\bibnamefont {Fini}},\ and\ \bibinfo
  {author} {\bibfnamefont {M.}~\bibnamefont {Matera}},\ }\bibfield  {title}
  {\bibinfo {title} {Resonance fluorescence of a two-level atom driven by a
  short laser pulse: extension to the off-resonance excitation},\ }\href
  {https://doi.org/10.1088/0953-4075/21/2/008} {\bibfield  {journal} {\bibinfo
  {journal} {J. Phys. B: At. Mol. Opt. Phys.}\ }\textbf {\bibinfo {volume}
  {21}},\ \bibinfo {pages} {239} (\bibinfo {year} {1988})}\BibitemShut
  {NoStop}%
\bibitem [{\citenamefont {Boos}\ \emph {et~al.}(2023)\citenamefont {Boos},
  \citenamefont {Kim}, \citenamefont {Bracht}, \citenamefont {Sbresny},
  \citenamefont {Kaspari}, \citenamefont {Cygorek}, \citenamefont {Riedl},
  \citenamefont {Bopp}, \citenamefont {Rauhaus}, \citenamefont {Calcagno},
  \citenamefont {Finley}, \citenamefont {Reiter},\ and\ \citenamefont
  {Mueller}}]{boos2023signatures}%
  \BibitemOpen
  \bibfield  {author} {\bibinfo {author} {\bibfnamefont {K.}~\bibnamefont
  {Boos}}, \bibinfo {author} {\bibfnamefont {S.~K.}\ \bibnamefont {Kim}},
  \bibinfo {author} {\bibfnamefont {T.}~\bibnamefont {Bracht}}, \bibinfo
  {author} {\bibfnamefont {F.}~\bibnamefont {Sbresny}}, \bibinfo {author}
  {\bibfnamefont {J.}~\bibnamefont {Kaspari}}, \bibinfo {author} {\bibfnamefont
  {M.}~\bibnamefont {Cygorek}}, \bibinfo {author} {\bibfnamefont
  {H.}~\bibnamefont {Riedl}}, \bibinfo {author} {\bibfnamefont {F.~W.}\
  \bibnamefont {Bopp}}, \bibinfo {author} {\bibfnamefont {W.}~\bibnamefont
  {Rauhaus}}, \bibinfo {author} {\bibfnamefont {C.}~\bibnamefont {Calcagno}},
  \bibinfo {author} {\bibfnamefont {J.~J.}\ \bibnamefont {Finley}}, \bibinfo
  {author} {\bibfnamefont {D.~E.}\ \bibnamefont {Reiter}},\ and\ \bibinfo
  {author} {\bibfnamefont {K.}~\bibnamefont {Mueller}},\ }\bibfield  {title}
  {\bibinfo {title} {Signatures of dynamically dressed states},\ }\href
  {https://arxiv.org/abs/2305.15827} {\bibfield  {journal} {\bibinfo  {journal}
  {arXiv preprint arXiv:2305.15827}\ } (\bibinfo {year} {2023})}\BibitemShut
  {NoStop}%
\bibitem [{\citenamefont {Liu}\ \emph {et~al.}(2023)\citenamefont {Liu},
  \citenamefont {Gustin}, \citenamefont {Liu}, \citenamefont {Li},
  \citenamefont {Yu}, \citenamefont {Ni}, \citenamefont {Niu}, \citenamefont
  {Hughes}, \citenamefont {Wang},\ and\ \citenamefont {Liu}}]{liu2023dynamic}%
  \BibitemOpen
  \bibfield  {author} {\bibinfo {author} {\bibfnamefont {S.}~\bibnamefont
  {Liu}}, \bibinfo {author} {\bibfnamefont {C.}~\bibnamefont {Gustin}},
  \bibinfo {author} {\bibfnamefont {H.}~\bibnamefont {Liu}}, \bibinfo {author}
  {\bibfnamefont {X.}~\bibnamefont {Li}}, \bibinfo {author} {\bibfnamefont
  {Y.}~\bibnamefont {Yu}}, \bibinfo {author} {\bibfnamefont {H.}~\bibnamefont
  {Ni}}, \bibinfo {author} {\bibfnamefont {Z.}~\bibnamefont {Niu}}, \bibinfo
  {author} {\bibfnamefont {S.}~\bibnamefont {Hughes}}, \bibinfo {author}
  {\bibfnamefont {X.}~\bibnamefont {Wang}},\ and\ \bibinfo {author}
  {\bibfnamefont {J.}~\bibnamefont {Liu}},\ }\bibfield  {title} {\bibinfo
  {title} {Dynamic resonance fluorescence in solid-state cavity quantum
  electrodynamics},\ }\href {https://arxiv.org/abs/2305.18776} {\bibfield
  {journal} {\bibinfo  {journal} {arXiv preprint arXiv:2305.18776}\ } (\bibinfo
  {year} {2023})}\BibitemShut {NoStop}%
\bibitem [{\citenamefont {Axt}\ and\ \citenamefont
  {Kuhn}(2004)}]{axt2004femtosecond}%
  \BibitemOpen
  \bibfield  {author} {\bibinfo {author} {\bibfnamefont {V.~M.}\ \bibnamefont
  {Axt}}\ and\ \bibinfo {author} {\bibfnamefont {T.}~\bibnamefont {Kuhn}},\
  }\bibfield  {title} {\bibinfo {title} {Femtosecond spectroscopy in
  semiconductors: a key to coherences, correlations and quantum kinetics},\
  }\href {https://iopscience.iop.org/article/10.1088/0034-4885/67/4/R01}
  {\bibfield  {journal} {\bibinfo  {journal} {Rep. Prog. Phys.}\ }\textbf
  {\bibinfo {volume} {67}},\ \bibinfo {pages} {433} (\bibinfo {year}
  {2004})}\BibitemShut {NoStop}%
\bibitem [{\citenamefont {Danckwerts}\ \emph {et~al.}(2006)\citenamefont
  {Danckwerts}, \citenamefont {Ahn}, \citenamefont {F{\"o}rstner},\ and\
  \citenamefont {Knorr}}]{danckwerts2006theory}%
  \BibitemOpen
  \bibfield  {author} {\bibinfo {author} {\bibfnamefont {J.}~\bibnamefont
  {Danckwerts}}, \bibinfo {author} {\bibfnamefont {K.}~\bibnamefont {Ahn}},
  \bibinfo {author} {\bibfnamefont {J.}~\bibnamefont {F{\"o}rstner}},\ and\
  \bibinfo {author} {\bibfnamefont {A.}~\bibnamefont {Knorr}},\ }\bibfield
  {title} {\bibinfo {title} {Theory of ultrafast nonlinear optics of
  {Coulomb-coupled} semiconductor quantum dots: {Rabi} oscillations and
  pump-probe spectra},\ }\href
  {https://journals.aps.org/prb/abstract/10.1103/PhysRevB.73.165318} {\bibfield
   {journal} {\bibinfo  {journal} {Phys. Rev. B}\ }\textbf {\bibinfo {volume}
  {73}},\ \bibinfo {pages} {165318} (\bibinfo {year} {2006})}\BibitemShut
  {NoStop}%
\bibitem [{\citenamefont {Sotier}\ \emph {et~al.}(2009)\citenamefont {Sotier},
  \citenamefont {Thomay}, \citenamefont {Hanke}, \citenamefont {Korger},
  \citenamefont {Mahapatra}, \citenamefont {Frey}, \citenamefont {Brunner},
  \citenamefont {Bratschitsch},\ and\ \citenamefont
  {Leitenstorfer}}]{sotier2009femtosecond}%
  \BibitemOpen
  \bibfield  {author} {\bibinfo {author} {\bibfnamefont {F.}~\bibnamefont
  {Sotier}}, \bibinfo {author} {\bibfnamefont {T.}~\bibnamefont {Thomay}},
  \bibinfo {author} {\bibfnamefont {T.}~\bibnamefont {Hanke}}, \bibinfo
  {author} {\bibfnamefont {J.}~\bibnamefont {Korger}}, \bibinfo {author}
  {\bibfnamefont {S.}~\bibnamefont {Mahapatra}}, \bibinfo {author}
  {\bibfnamefont {A.}~\bibnamefont {Frey}}, \bibinfo {author} {\bibfnamefont
  {K.}~\bibnamefont {Brunner}}, \bibinfo {author} {\bibfnamefont
  {R.}~\bibnamefont {Bratschitsch}},\ and\ \bibinfo {author} {\bibfnamefont
  {A.}~\bibnamefont {Leitenstorfer}},\ }\bibfield  {title} {\bibinfo {title}
  {Femtosecond few-fermion dynamics and deterministic single-photon gain in a
  quantum dot},\ }\href {https://www.nature.com/articles/nphys1229} {\bibfield
  {journal} {\bibinfo  {journal} {Nat. Phys.}\ }\textbf {\bibinfo {volume}
  {5}},\ \bibinfo {pages} {352} (\bibinfo {year} {2009})}\BibitemShut {NoStop}%
\bibitem [{\citenamefont {Henzler}\ \emph {et~al.}(2021)\citenamefont
  {Henzler}, \citenamefont {Traum}, \citenamefont {Holtkemper}, \citenamefont
  {Nabben}, \citenamefont {Erbe}, \citenamefont {Reiter}, \citenamefont {Kuhn},
  \citenamefont {Mahapatra}, \citenamefont {Brunner}, \citenamefont {Seletskiy}
  \emph {et~al.}}]{henzler2021femtosecond}%
  \BibitemOpen
  \bibfield  {author} {\bibinfo {author} {\bibfnamefont {P.}~\bibnamefont
  {Henzler}}, \bibinfo {author} {\bibfnamefont {C.}~\bibnamefont {Traum}},
  \bibinfo {author} {\bibfnamefont {M.}~\bibnamefont {Holtkemper}}, \bibinfo
  {author} {\bibfnamefont {D.}~\bibnamefont {Nabben}}, \bibinfo {author}
  {\bibfnamefont {M.}~\bibnamefont {Erbe}}, \bibinfo {author} {\bibfnamefont
  {D.~E.}\ \bibnamefont {Reiter}}, \bibinfo {author} {\bibfnamefont
  {T.}~\bibnamefont {Kuhn}}, \bibinfo {author} {\bibfnamefont {S.}~\bibnamefont
  {Mahapatra}}, \bibinfo {author} {\bibfnamefont {K.}~\bibnamefont {Brunner}},
  \bibinfo {author} {\bibfnamefont {D.~V.}\ \bibnamefont {Seletskiy}}, \emph
  {et~al.},\ }\bibfield  {title} {\bibinfo {title} {Femtosecond transfer and
  manipulation of persistent hot-trion coherence in a single {CdSe/ZnSe}
  quantum dot},\ }\href
  {https://journals.aps.org/prl/abstract/10.1103/PhysRevLett.126.067402}
  {\bibfield  {journal} {\bibinfo  {journal} {Phys. Rev. Lett.}\ }\textbf
  {\bibinfo {volume} {126}},\ \bibinfo {pages} {067402} (\bibinfo {year}
  {2021})}\BibitemShut {NoStop}%
\bibitem [{\citenamefont {Zecherle}\ \emph {et~al.}(2010)\citenamefont
  {Zecherle}, \citenamefont {Ruppert}, \citenamefont {Clark}, \citenamefont
  {Abstreiter}, \citenamefont {Finley},\ and\ \citenamefont
  {Betz}}]{zecherle2010ultrafast}%
  \BibitemOpen
  \bibfield  {author} {\bibinfo {author} {\bibfnamefont {M.}~\bibnamefont
  {Zecherle}}, \bibinfo {author} {\bibfnamefont {C.}~\bibnamefont {Ruppert}},
  \bibinfo {author} {\bibfnamefont {E.~C.}\ \bibnamefont {Clark}}, \bibinfo
  {author} {\bibfnamefont {G.}~\bibnamefont {Abstreiter}}, \bibinfo {author}
  {\bibfnamefont {J.~J.}\ \bibnamefont {Finley}},\ and\ \bibinfo {author}
  {\bibfnamefont {M.}~\bibnamefont {Betz}},\ }\bibfield  {title} {\bibinfo
  {title} {Ultrafast few-fermion optoelectronics in a single self-assembled
  {InGaAs/GaAs} quantum dot},\ }\href
  {https://journals.aps.org/prb/abstract/10.1103/PhysRevB.82.125314} {\bibfield
   {journal} {\bibinfo  {journal} {Phys. Rev. B}\ }\textbf {\bibinfo {volume}
  {82}},\ \bibinfo {pages} {125314} (\bibinfo {year} {2010})}\BibitemShut
  {NoStop}%
\bibitem [{\citenamefont {Suzuki}\ \emph {et~al.}(2016)\citenamefont {Suzuki},
  \citenamefont {Singh}, \citenamefont {Bayer}, \citenamefont {Ludwig},
  \citenamefont {Wieck},\ and\ \citenamefont {Cundiff}}]{suzuki2016coherent}%
  \BibitemOpen
  \bibfield  {author} {\bibinfo {author} {\bibfnamefont {T.}~\bibnamefont
  {Suzuki}}, \bibinfo {author} {\bibfnamefont {R.}~\bibnamefont {Singh}},
  \bibinfo {author} {\bibfnamefont {M.}~\bibnamefont {Bayer}}, \bibinfo
  {author} {\bibfnamefont {A.}~\bibnamefont {Ludwig}}, \bibinfo {author}
  {\bibfnamefont {A.~D.}\ \bibnamefont {Wieck}},\ and\ \bibinfo {author}
  {\bibfnamefont {S.~T.}\ \bibnamefont {Cundiff}},\ }\bibfield  {title}
  {\bibinfo {title} {Coherent control of the exciton-biexciton system in an
  {InAs} self-assembled quantum dot ensemble},\ }\href
  {https://journals.aps.org/prl/abstract/10.1103/PhysRevLett.117.157402}
  {\bibfield  {journal} {\bibinfo  {journal} {Phys. Rev. Lett}\ }\textbf
  {\bibinfo {volume} {117}},\ \bibinfo {pages} {157402} (\bibinfo {year}
  {2016})}\BibitemShut {NoStop}%
\bibitem [{\citenamefont {Fras}\ \emph {et~al.}(2016)\citenamefont {Fras},
  \citenamefont {Mermillod}, \citenamefont {Nogues}, \citenamefont {Hoarau},
  \citenamefont {Schneider}, \citenamefont {Kamp}, \citenamefont {H{\"o}fling},
  \citenamefont {Langbein},\ and\ \citenamefont {Kasprzak}}]{fras2016multi}%
  \BibitemOpen
  \bibfield  {author} {\bibinfo {author} {\bibfnamefont {F.}~\bibnamefont
  {Fras}}, \bibinfo {author} {\bibfnamefont {Q.}~\bibnamefont {Mermillod}},
  \bibinfo {author} {\bibfnamefont {G.}~\bibnamefont {Nogues}}, \bibinfo
  {author} {\bibfnamefont {C.}~\bibnamefont {Hoarau}}, \bibinfo {author}
  {\bibfnamefont {C.}~\bibnamefont {Schneider}}, \bibinfo {author}
  {\bibfnamefont {M.}~\bibnamefont {Kamp}}, \bibinfo {author} {\bibfnamefont
  {S.}~\bibnamefont {H{\"o}fling}}, \bibinfo {author} {\bibfnamefont
  {W.}~\bibnamefont {Langbein}},\ and\ \bibinfo {author} {\bibfnamefont
  {J.}~\bibnamefont {Kasprzak}},\ }\bibfield  {title} {\bibinfo {title}
  {Multi-wave coherent control of a solid-state single emitter},\ }\href
  {https://www.nature.com/articles/nphoton.2016.2} {\bibfield  {journal}
  {\bibinfo  {journal} {Nature Photon.}\ }\textbf {\bibinfo {volume} {10}},\
  \bibinfo {pages} {155} (\bibinfo {year} {2016})}\BibitemShut {NoStop}%
\bibitem [{\citenamefont {Richter}\ \emph {et~al.}(2018)\citenamefont
  {Richter}, \citenamefont {Singh}, \citenamefont {Siemens},\ and\
  \citenamefont {Cundiff}}]{richter2018deconvolution}%
  \BibitemOpen
  \bibfield  {author} {\bibinfo {author} {\bibfnamefont {M.}~\bibnamefont
  {Richter}}, \bibinfo {author} {\bibfnamefont {R.}~\bibnamefont {Singh}},
  \bibinfo {author} {\bibfnamefont {M.}~\bibnamefont {Siemens}},\ and\ \bibinfo
  {author} {\bibfnamefont {S.~T.}\ \bibnamefont {Cundiff}},\ }\bibfield
  {title} {\bibinfo {title} {Deconvolution of optical multidimensional coherent
  spectra},\ }\href {https://www.science.org/doi/full/10.1126/sciadv.aar7697}
  {\bibfield  {journal} {\bibinfo  {journal} {Science Advances}\ }\textbf
  {\bibinfo {volume} {4}},\ \bibinfo {pages} {eaar7697} (\bibinfo {year}
  {2018})}\BibitemShut {NoStop}%
\bibitem [{\citenamefont {Grisard}\ \emph {et~al.}(2022)\citenamefont
  {Grisard}, \citenamefont {Rose}, \citenamefont {Trifonov}, \citenamefont
  {Reichhardt}, \citenamefont {Reiter}, \citenamefont {Reichelt}, \citenamefont
  {Schneider}, \citenamefont {Kamp}, \citenamefont {H{\"o}fling}, \citenamefont
  {Bayer} \emph {et~al.}}]{grisard2022multiple}%
  \BibitemOpen
  \bibfield  {author} {\bibinfo {author} {\bibfnamefont {S.}~\bibnamefont
  {Grisard}}, \bibinfo {author} {\bibfnamefont {H.}~\bibnamefont {Rose}},
  \bibinfo {author} {\bibfnamefont {A.~V.}\ \bibnamefont {Trifonov}}, \bibinfo
  {author} {\bibfnamefont {R.}~\bibnamefont {Reichhardt}}, \bibinfo {author}
  {\bibfnamefont {D.~E.}\ \bibnamefont {Reiter}}, \bibinfo {author}
  {\bibfnamefont {M.}~\bibnamefont {Reichelt}}, \bibinfo {author}
  {\bibfnamefont {C.}~\bibnamefont {Schneider}}, \bibinfo {author}
  {\bibfnamefont {M.}~\bibnamefont {Kamp}}, \bibinfo {author} {\bibfnamefont
  {S.}~\bibnamefont {H{\"o}fling}}, \bibinfo {author} {\bibfnamefont
  {M.}~\bibnamefont {Bayer}}, \emph {et~al.},\ }\bibfield  {title} {\bibinfo
  {title} {Multiple {Rabi} rotations of trions in {InGaAs} quantum dots
  observed by photon echo spectroscopy with spatially shaped laser pulses},\
  }\href {https://link.aps.org/doi/10.1103/PhysRevB.106.205408} {\bibfield
  {journal} {\bibinfo  {journal} {Phys. Rev. B}\ }\textbf {\bibinfo {volume}
  {106}},\ \bibinfo {pages} {205408} (\bibinfo {year} {2022})}\BibitemShut
  {NoStop}%
\bibitem [{\citenamefont {Ruppert}\ \emph {et~al.}(2012)\citenamefont
  {Ruppert}, \citenamefont {Lohrenz}, \citenamefont {Thunich},\ and\
  \citenamefont {Betz}}]{betz2012}%
  \BibitemOpen
  \bibfield  {author} {\bibinfo {author} {\bibfnamefont {C.}~\bibnamefont
  {Ruppert}}, \bibinfo {author} {\bibfnamefont {J.}~\bibnamefont {Lohrenz}},
  \bibinfo {author} {\bibfnamefont {S.}~\bibnamefont {Thunich}},\ and\ \bibinfo
  {author} {\bibfnamefont {M.}~\bibnamefont {Betz}},\ }\bibfield  {title}
  {\bibinfo {title} {Ultrafast field-resolved semiconductor spectroscopy
  utilizing quantum interference control of currents},\ }\href
  {https://doi.org/10.1364/OL.37.003879} {\bibfield  {journal} {\bibinfo
  {journal} {Opt. Lett.}\ }\textbf {\bibinfo {volume} {37}},\ \bibinfo {pages}
  {3879} (\bibinfo {year} {2012})}\BibitemShut {NoStop}%
\bibitem [{\citenamefont {Guenther}\ \emph {et~al.}(2002)\citenamefont
  {Guenther}, \citenamefont {Lienau}, \citenamefont {Elsaesser}, \citenamefont
  {Glanemann}, \citenamefont {Axt}, \citenamefont {Kuhn}, \citenamefont
  {Eshlaghi},\ and\ \citenamefont {Wieck}}]{guenther2002}%
  \BibitemOpen
  \bibfield  {author} {\bibinfo {author} {\bibfnamefont {T.}~\bibnamefont
  {Guenther}}, \bibinfo {author} {\bibfnamefont {C.}~\bibnamefont {Lienau}},
  \bibinfo {author} {\bibfnamefont {T.}~\bibnamefont {Elsaesser}}, \bibinfo
  {author} {\bibfnamefont {M.}~\bibnamefont {Glanemann}}, \bibinfo {author}
  {\bibfnamefont {V.~M.}\ \bibnamefont {Axt}}, \bibinfo {author} {\bibfnamefont
  {T.}~\bibnamefont {Kuhn}}, \bibinfo {author} {\bibfnamefont {S.}~\bibnamefont
  {Eshlaghi}},\ and\ \bibinfo {author} {\bibfnamefont {A.~D.}\ \bibnamefont
  {Wieck}},\ }\bibfield  {title} {\bibinfo {title} {Coherent nonlinear optical
  response of single quantum dots studied by ultrafast near-field
  spectroscopy},\ }\href
  {https://link.aps.org/doi/10.1103/PhysRevLett.89.057401} {\bibfield
  {journal} {\bibinfo  {journal} {Phys. Rev. Lett.}\ }\textbf {\bibinfo
  {volume} {89}},\ \bibinfo {pages} {057401} (\bibinfo {year}
  {2002})}\BibitemShut {NoStop}%
\bibitem [{\citenamefont {Yan}\ \emph {et~al.}(2011)\citenamefont {Yan},
  \citenamefont {Seidel},\ and\ \citenamefont {Tan}}]{yan2011}%
  \BibitemOpen
  \bibfield  {author} {\bibinfo {author} {\bibfnamefont {S.}~\bibnamefont
  {Yan}}, \bibinfo {author} {\bibfnamefont {M.~T.}\ \bibnamefont {Seidel}},\
  and\ \bibinfo {author} {\bibfnamefont {H.-S.}\ \bibnamefont {Tan}},\
  }\bibfield  {title} {\bibinfo {title} {Perturbed free induction decay in
  ultrafast {mid-IR} pump–probe spectroscopy},\ }\href
  {https://www.sciencedirect.com/science/article/pii/S0009261411012656}
  {\bibfield  {journal} {\bibinfo  {journal} {Chem. Phys. Lett.}\ }\textbf
  {\bibinfo {volume} {517}},\ \bibinfo {pages} {36} (\bibinfo {year}
  {2011})}\BibitemShut {NoStop}%
\bibitem [{\citenamefont {Nuernberger}\ \emph {et~al.}(2009)\citenamefont
  {Nuernberger}, \citenamefont {Lee}, \citenamefont {Bonvalet}, \citenamefont
  {Polack}, \citenamefont {Vos}, \citenamefont {Alexandrou},\ and\
  \citenamefont {Joffre}}]{nuernberger2009}%
  \BibitemOpen
  \bibfield  {author} {\bibinfo {author} {\bibfnamefont {P.}~\bibnamefont
  {Nuernberger}}, \bibinfo {author} {\bibfnamefont {K.~F.}\ \bibnamefont
  {Lee}}, \bibinfo {author} {\bibfnamefont {A.}~\bibnamefont {Bonvalet}},
  \bibinfo {author} {\bibfnamefont {T.}~\bibnamefont {Polack}}, \bibinfo
  {author} {\bibfnamefont {M.~H.}\ \bibnamefont {Vos}}, \bibinfo {author}
  {\bibfnamefont {A.}~\bibnamefont {Alexandrou}},\ and\ \bibinfo {author}
  {\bibfnamefont {M.}~\bibnamefont {Joffre}},\ }\bibfield  {title} {\bibinfo
  {title} {Suppression of perturbed free-induction decay and noise in
  experimental ultrafast pump-probe data},\ }\href
  {https://opg.optica.org/ol/abstract.cfm?URI=ol-34-20-3226} {\bibfield
  {journal} {\bibinfo  {journal} {Opt. Lett.}\ }\textbf {\bibinfo {volume}
  {34}},\ \bibinfo {pages} {3226} (\bibinfo {year} {2009})}\BibitemShut
  {NoStop}%
\bibitem [{\citenamefont {Mondal}\ \emph {et~al.}(2018)\citenamefont {Mondal},
  \citenamefont {Roy}, \citenamefont {Pal},\ and\ \citenamefont
  {Bansal}}]{mondal2018}%
  \BibitemOpen
  \bibfield  {author} {\bibinfo {author} {\bibfnamefont {R.}~\bibnamefont
  {Mondal}}, \bibinfo {author} {\bibfnamefont {B.}~\bibnamefont {Roy}},
  \bibinfo {author} {\bibfnamefont {B.}~\bibnamefont {Pal}},\ and\ \bibinfo
  {author} {\bibfnamefont {B.}~\bibnamefont {Bansal}},\ }\bibfield  {title}
  {\bibinfo {title} {How pump–probe differential reflectivity at negative
  delay yields the perturbed-free-induction-decay: theory of the experiment and
  its verification},\ }\href {https://dx.doi.org/10.1088/1361-648X/aaed79}
  {\bibfield  {journal} {\bibinfo  {journal} {J. Phys. Condens. Matter}\
  }\textbf {\bibinfo {volume} {30}},\ \bibinfo {pages} {505902} (\bibinfo
  {year} {2018})}\BibitemShut {NoStop}%
\bibitem [{\citenamefont {Murotani}\ \emph {et~al.}(2018)\citenamefont
  {Murotani}, \citenamefont {Takayama}, \citenamefont {Sekiguchi},
  \citenamefont {Kim}, \citenamefont {Akiyama},\ and\ \citenamefont
  {Shimano}}]{murotani2018}%
  \BibitemOpen
  \bibfield  {author} {\bibinfo {author} {\bibfnamefont {Y.}~\bibnamefont
  {Murotani}}, \bibinfo {author} {\bibfnamefont {M.}~\bibnamefont {Takayama}},
  \bibinfo {author} {\bibfnamefont {F.}~\bibnamefont {Sekiguchi}}, \bibinfo
  {author} {\bibfnamefont {C.}~\bibnamefont {Kim}}, \bibinfo {author}
  {\bibfnamefont {H.}~\bibnamefont {Akiyama}},\ and\ \bibinfo {author}
  {\bibfnamefont {R.}~\bibnamefont {Shimano}},\ }\bibfield  {title} {\bibinfo
  {title} {Terahertz field-induced ionization and perturbed free induction
  decay of excitons in bulk {GaAs}},\ }\href
  {https://dx.doi.org/10.1088/1361-6463/aaa989} {\bibfield  {journal} {\bibinfo
   {journal} {J. Phys. D: Appl. Phys.}\ }\textbf {\bibinfo {volume} {51}},\
  \bibinfo {pages} {114001} (\bibinfo {year} {2018})}\BibitemShut {NoStop}%
\bibitem [{\citenamefont {Brosseau}\ \emph {et~al.}(2023)\citenamefont
  {Brosseau}, \citenamefont {Seiler}, \citenamefont {Palato}, \citenamefont
  {Sonnichsen}, \citenamefont {Baker}, \citenamefont {Socie}, \citenamefont
  {Strandell},\ and\ \citenamefont {Kambhampati}}]{brosseau2023}%
  \BibitemOpen
  \bibfield  {author} {\bibinfo {author} {\bibfnamefont {P.}~\bibnamefont
  {Brosseau}}, \bibinfo {author} {\bibfnamefont {H.}~\bibnamefont {Seiler}},
  \bibinfo {author} {\bibfnamefont {S.}~\bibnamefont {Palato}}, \bibinfo
  {author} {\bibfnamefont {C.}~\bibnamefont {Sonnichsen}}, \bibinfo {author}
  {\bibfnamefont {H.}~\bibnamefont {Baker}}, \bibinfo {author} {\bibfnamefont
  {E.}~\bibnamefont {Socie}}, \bibinfo {author} {\bibfnamefont
  {D.}~\bibnamefont {Strandell}},\ and\ \bibinfo {author} {\bibfnamefont
  {P.}~\bibnamefont {Kambhampati}},\ }\bibfield  {title} {\bibinfo {title}
  {Perturbed free induction decay obscures early time dynamics in
  two-dimensional electronic spectroscopy: The case of semiconductor
  nanocrystals},\ }\href {https://doi.org/10.1063/5.0138252} {\bibfield
  {journal} {\bibinfo  {journal} {J. Chem. Phys.}\ }\textbf {\bibinfo {volume}
  {158}},\ \bibinfo {pages} {084201} (\bibinfo {year} {2023})}\BibitemShut
  {NoStop}%
\bibitem [{\citenamefont {Hamm}(1995)}]{hamm1995}%
  \BibitemOpen
  \bibfield  {author} {\bibinfo {author} {\bibfnamefont {P.}~\bibnamefont
  {Hamm}},\ }\bibfield  {title} {\bibinfo {title} {{Coherent effects in
  femtosecond infrared spectroscopy}},\ }\href
  {https://doi.org/10.1016/0301-0104(95)00262-6} {\bibfield  {journal}
  {\bibinfo  {journal} {J. Chem. Phys.}\ }\textbf {\bibinfo {volume} {200}},\
  \bibinfo {pages} {415} (\bibinfo {year} {1995})}\BibitemShut {NoStop}%
\bibitem [{\citenamefont {Seidner}\ \emph {et~al.}(1995)\citenamefont
  {Seidner}, \citenamefont {Stock},\ and\ \citenamefont
  {Domcke}}]{seidner1995}%
  \BibitemOpen
  \bibfield  {author} {\bibinfo {author} {\bibfnamefont {L.}~\bibnamefont
  {Seidner}}, \bibinfo {author} {\bibfnamefont {G.}~\bibnamefont {Stock}},\
  and\ \bibinfo {author} {\bibfnamefont {W.}~\bibnamefont {Domcke}},\
  }\bibfield  {title} {\bibinfo {title} {Nonperturbative approach to
  femtosecond spectroscopy: General theory and application to multidimensional
  nonadiabatic photoisomerization processes},\ }\href
  {https://doi.org/10.1063/1.469586} {\bibfield  {journal} {\bibinfo  {journal}
  {J. Chem. Phys.}\ }\textbf {\bibinfo {volume} {103}},\ \bibinfo {pages}
  {3998} (\bibinfo {year} {1995})}\BibitemShut {NoStop}%
\bibitem [{\citenamefont {Wolpert}\ \emph {et~al.}(2012)\citenamefont
  {Wolpert}, \citenamefont {Dicken}, \citenamefont {Atkinson}, \citenamefont
  {Wang}, \citenamefont {Rastelli}, \citenamefont {Schmidt}, \citenamefont
  {Giessen},\ and\ \citenamefont {Lippitz}}]{wolpert2012}%
  \BibitemOpen
  \bibfield  {author} {\bibinfo {author} {\bibfnamefont {C.}~\bibnamefont
  {Wolpert}}, \bibinfo {author} {\bibfnamefont {C.}~\bibnamefont {Dicken}},
  \bibinfo {author} {\bibfnamefont {P.}~\bibnamefont {Atkinson}}, \bibinfo
  {author} {\bibfnamefont {L.}~\bibnamefont {Wang}}, \bibinfo {author}
  {\bibfnamefont {A.}~\bibnamefont {Rastelli}}, \bibinfo {author}
  {\bibfnamefont {O.~G.}\ \bibnamefont {Schmidt}}, \bibinfo {author}
  {\bibfnamefont {H.}~\bibnamefont {Giessen}},\ and\ \bibinfo {author}
  {\bibfnamefont {M.}~\bibnamefont {Lippitz}},\ }\bibfield  {title} {\bibinfo
  {title} {Transient reflection: A versatile technique for ultrafast
  spectroscopy of a single quantum dot in complex environments},\ }\href
  {https://doi.org/10.1021/nl203804n} {\bibfield  {journal} {\bibinfo
  {journal} {Nano Lett.}\ }\textbf {\bibinfo {volume} {12}},\ \bibinfo {pages}
  {453} (\bibinfo {year} {2012})}\BibitemShut {NoStop}%
\bibitem [{\citenamefont {Reiter}(2017)}]{Reiter2017}%
  \BibitemOpen
  \bibfield  {author} {\bibinfo {author} {\bibfnamefont {D.~E.}\ \bibnamefont
  {Reiter}},\ }\bibfield  {title} {\bibinfo {title} {Time-resolved pump-probe
  signals of a continuously driven quantum dot affected by phonons},\ }\href
  {https://doi.org/10.1103/PhysRevB.95.125308} {\bibfield  {journal} {\bibinfo
  {journal} {Phys. Rev. B}\ }\textbf {\bibinfo {volume} {95}},\ \bibinfo
  {pages} {125308} (\bibinfo {year} {2017})}\BibitemShut {NoStop}%
\bibitem [{\citenamefont {Kla{\ss}en}\ and\ \citenamefont
  {Reiter}(2021)}]{klassen2021optical}%
  \BibitemOpen
  \bibfield  {author} {\bibinfo {author} {\bibfnamefont {M.~R.}\ \bibnamefont
  {Kla{\ss}en}}\ and\ \bibinfo {author} {\bibfnamefont {D.~E.}\ \bibnamefont
  {Reiter}},\ }\bibfield  {title} {\bibinfo {title} {Optical signals to monitor
  the dynamics of phonon-modified {Rabi} oscillations in a quantum dot},\
  }\href {https://onlinelibrary.wiley.com/doi/full/10.1002/andp.202100086}
  {\bibfield  {journal} {\bibinfo  {journal} {Ann. Phys.}\ }\textbf {\bibinfo
  {volume} {533}},\ \bibinfo {pages} {2100086} (\bibinfo {year}
  {2021})}\BibitemShut {NoStop}%
\bibitem [{\citenamefont {Besombes}\ \emph {et~al.}(2001)\citenamefont
  {Besombes}, \citenamefont {Kheng}, \citenamefont {Marsal},\ and\
  \citenamefont {Mariette}}]{Besombes2001}%
  \BibitemOpen
  \bibfield  {author} {\bibinfo {author} {\bibfnamefont {L.}~\bibnamefont
  {Besombes}}, \bibinfo {author} {\bibfnamefont {K.}~\bibnamefont {Kheng}},
  \bibinfo {author} {\bibfnamefont {L.}~\bibnamefont {Marsal}},\ and\ \bibinfo
  {author} {\bibfnamefont {H.}~\bibnamefont {Mariette}},\ }\bibfield  {title}
  {\bibinfo {title} {Acoustic phonon broadening mechanism in single quantum dot
  emission},\ }\href {https://doi.org/10.1103/PhysRevB.63.155307} {\bibfield
  {journal} {\bibinfo  {journal} {Phys. Rev. B}\ }\textbf {\bibinfo {volume}
  {63}},\ \bibinfo {pages} {155307} (\bibinfo {year} {2001})}\BibitemShut
  {NoStop}%
\bibitem [{\citenamefont {Krummheuer}\ \emph {et~al.}(2002)\citenamefont
  {Krummheuer}, \citenamefont {Axt},\ and\ \citenamefont
  {Kuhn}}]{Krummheuer2002}%
  \BibitemOpen
  \bibfield  {author} {\bibinfo {author} {\bibfnamefont {B.}~\bibnamefont
  {Krummheuer}}, \bibinfo {author} {\bibfnamefont {V.~M.}\ \bibnamefont
  {Axt}},\ and\ \bibinfo {author} {\bibfnamefont {T.}~\bibnamefont {Kuhn}},\
  }\bibfield  {title} {\bibinfo {title} {Theory of pure dephasing and the
  resulting absorption line shape in semiconductor quantum dots},\ }\href
  {https://doi.org/10.1103/PhysRevB.65.195313} {\bibfield  {journal} {\bibinfo
  {journal} {Phys. Rev. B}\ }\textbf {\bibinfo {volume} {65}},\ \bibinfo
  {pages} {195313} (\bibinfo {year} {2002})}\BibitemShut {NoStop}%
\bibitem [{\citenamefont {Kr{\"u}gel}\ \emph {et~al.}(2005)\citenamefont
  {Kr{\"u}gel}, \citenamefont {Axt}, \citenamefont {Kuhn}, \citenamefont
  {Machnikowski},\ and\ \citenamefont {Vagov}}]{krugel2005}%
  \BibitemOpen
  \bibfield  {author} {\bibinfo {author} {\bibfnamefont {A.}~\bibnamefont
  {Kr{\"u}gel}}, \bibinfo {author} {\bibfnamefont {V.~M.}\ \bibnamefont {Axt}},
  \bibinfo {author} {\bibfnamefont {T.}~\bibnamefont {Kuhn}}, \bibinfo {author}
  {\bibfnamefont {P.}~\bibnamefont {Machnikowski}},\ and\ \bibinfo {author}
  {\bibfnamefont {A.}~\bibnamefont {Vagov}},\ }\bibfield  {title} {\bibinfo
  {title} {The role of acoustic phonons for {Rabi} oscillations in
  semiconductor quantum dots},\ }\href
  {https://doi.org/10.1007/s00340-005-1984-1} {\bibfield  {journal} {\bibinfo
  {journal} {Appl. Phys. B}\ }\textbf {\bibinfo {volume} {81}},\ \bibinfo
  {pages} {897} (\bibinfo {year} {2005})}\BibitemShut {NoStop}%
\bibitem [{\citenamefont {Roy}\ and\ \citenamefont
  {Hughes}(2012)}]{roy2012polaron}%
  \BibitemOpen
  \bibfield  {author} {\bibinfo {author} {\bibfnamefont {C.}~\bibnamefont
  {Roy}}\ and\ \bibinfo {author} {\bibfnamefont {S.}~\bibnamefont {Hughes}},\
  }\bibfield  {title} {\bibinfo {title} {Polaron master equation theory of the
  quantum-dot {Mollow} triplet in a semiconductor {cavity-QED} system},\ }\href
  {https://journals.aps.org/prb/abstract/10.1103/PhysRevB.85.115309} {\bibfield
   {journal} {\bibinfo  {journal} {Phys. Rev. B}\ }\textbf {\bibinfo {volume}
  {85}},\ \bibinfo {pages} {115309} (\bibinfo {year} {2012})}\BibitemShut
  {NoStop}%
\bibitem [{\citenamefont {Cosacchi}\ \emph {et~al.}(2018)\citenamefont
  {Cosacchi}, \citenamefont {Cygorek}, \citenamefont {Ungar}, \citenamefont
  {Barth}, \citenamefont {Vagov},\ and\ \citenamefont
  {Axt}}]{cosacchi2018path}%
  \BibitemOpen
  \bibfield  {author} {\bibinfo {author} {\bibfnamefont {M.}~\bibnamefont
  {Cosacchi}}, \bibinfo {author} {\bibfnamefont {M.}~\bibnamefont {Cygorek}},
  \bibinfo {author} {\bibfnamefont {F.}~\bibnamefont {Ungar}}, \bibinfo
  {author} {\bibfnamefont {A.~M.}\ \bibnamefont {Barth}}, \bibinfo {author}
  {\bibfnamefont {A.}~\bibnamefont {Vagov}},\ and\ \bibinfo {author}
  {\bibfnamefont {V.~M.}\ \bibnamefont {Axt}},\ }\bibfield  {title} {\bibinfo
  {title} {Path-integral approach for nonequilibrium multitime correlation
  functions of open quantum systems coupled to {Markovian} and {non-Markovian}
  environments},\ }\href
  {https://journals.aps.org/prb/abstract/10.1103/PhysRevB.98.125302} {\bibfield
   {journal} {\bibinfo  {journal} {Phys. Rev. B}\ }\textbf {\bibinfo {volume}
  {98}},\ \bibinfo {pages} {125302} (\bibinfo {year} {2018})}\BibitemShut
  {NoStop}%
\bibitem [{\citenamefont {Cygorek}\ \emph {et~al.}(2022)\citenamefont
  {Cygorek}, \citenamefont {Cosacchi}, \citenamefont {Vagov}, \citenamefont
  {Axt}, \citenamefont {Lovett}, \citenamefont {Keeling},\ and\ \citenamefont
  {Gauger}}]{cygorek2022}%
  \BibitemOpen
  \bibfield  {author} {\bibinfo {author} {\bibfnamefont {M.}~\bibnamefont
  {Cygorek}}, \bibinfo {author} {\bibfnamefont {M.}~\bibnamefont {Cosacchi}},
  \bibinfo {author} {\bibfnamefont {A.}~\bibnamefont {Vagov}}, \bibinfo
  {author} {\bibfnamefont {V.~M.}\ \bibnamefont {Axt}}, \bibinfo {author}
  {\bibfnamefont {B.~W.}\ \bibnamefont {Lovett}}, \bibinfo {author}
  {\bibfnamefont {J.}~\bibnamefont {Keeling}},\ and\ \bibinfo {author}
  {\bibfnamefont {E.~M.}\ \bibnamefont {Gauger}},\ }\bibfield  {title}
  {\bibinfo {title} {Simulation of open quantum systems by automated
  compression of arbitrary environments},\ }\href
  {https://doi.org/10.1038/s41567-022-01544-9} {\bibfield  {journal} {\bibinfo
  {journal} {Nat. Phys.}\ }\textbf {\bibinfo {volume} {18}},\ \bibinfo {pages}
  {662} (\bibinfo {year} {2022})}\BibitemShut {NoStop}%
\end{thebibliography}%
\end{document}